\documentclass[12pt,a4paper]{article}
\pdfoutput=1

\usepackage{a4wide}

\usepackage{latexsym}
\usepackage{epsf}
\usepackage{amssymb}
\usepackage{graphicx}
\usepackage{amsmath, cite}
\usepackage{amsmath,amssymb,amsthm}
\usepackage{verbatim}
\usepackage{hyperref}
\usepackage{amsmath}
\usepackage{amsfonts}
\usepackage{xfrac}
\usepackage{slashed}
\usepackage{tikz}
\usepackage{cancel}
\usepackage{bbm}

\usepackage{mathtools}

\usepackage[utf8]{inputenc}

\usepackage{fancyhdr}
\usepackage{datetime}

\usepackage{makeidx}

\renewcommand{\Im}{\textrm{Im}}
\renewcommand{\d}{\textrm{d}}

\newcommand{\Tr}{\textrm{Tr}}
\newcommand{\STr}{\textrm{STr}}

\newcommand{\w}{\wedge}

\newcommand\varpm{\mathbin{\vcenter{\hbox{%
  \oalign{\hfil$\scriptstyle+$\hfil\cr
          \noalign{\kern-.3ex}
          $\scriptscriptstyle({-})$\cr}%
}}}}

\newcommand\varmp{\mathbin{\vcenter{\hbox{%
   \oalign{\hfil$\scriptstyle-$\hfil\cr
           \noalign{\kern-.3ex}
          $\scriptscriptstyle({+})$\cr}%
}}}}

\newcommand{\WZ}{\textrm{\tiny WZ}}
\newcommand{\DBI}{\textrm{\tiny DBI}}
\newcommand{\p}{\partial}

\usetikzlibrary{arrows,automata,positioning,calc,trees,decorations.pathmorphing,decorations.markings}

\DeclareMathOperator*{\tooo}{\longrightarrow}
\usepackage{fixfoot}

\begin{document}
\numberwithin{equation}{section}
\begin{flushright}
\small
 ROM2F/2019/04
\normalsize
\end{flushright}

\vspace{1 cm}

\begin{center}
{\LARGE \textbf{{The two faces of T-branes }}}

\vspace{2 cm} {\large Iosif Bena${}^1$, Johan Bl{\aa}b{\"a}ck${}^2$ , Raffaele Savelli${}^2$ , Gianluca Zoccarato${}^3$ }\\
\vspace{1 cm}
\small{
${}^1$
Institut de Physique Th\'eorique, Universit\'e Paris Saclay, CEA, CNRS,\\
Orme des Merisiers, F-91191 Gif sur Yvette, France\\[2mm]
${}^2$ Dipartimento di Fisica, Universit\`a di Roma ``Tor Vergata" \& INFN - Sezione di Roma2 \\
Via della Ricerca Scientifica 1, 00133 Roma, ITALY \\[2mm]
${}^3$ Department of Physics and Astronomy, University of Pennsylvania, \\ Philadelphia, PA 19104, USA
\\[8mm]}


\vspace{0.1cm} {\small\upshape\ttfamily ${}^1$ iosif.bena [AT] ipht.fr \\ ${}^2$ \{johan.blaback, raffaele.savelli\} [AT] roma2.infn.it \\ ${}^3$ gzoc [AT] sas.upenn.edu} \\


\vspace{2.3 cm}

\textbf{Abstract}
\end{center}
We establish a brane-brane duality connecting T-branes to collections of ordinary D-branes. T-branes are intrinsically non-Abelian brane configurations with worldvolume flux, whereas their duals consist of Abelian brane systems that encode the T-brane data in their curvature. We argue that the new Abelian picture provides a reliable description of T-branes when their non-Abelian fields have large expectation values in string units. To confirm this duality, we match the energy density and all the electromagnetic couplings on both sides. A key step in this derivation is a non-trivial factorization of the symmetrized-trace non-Abelian Dirac-Born-Infeld action when evaluated on solutions of the $\alpha'$-corrected Hitchin system.

\begin{quotation}

\end{quotation}

\thispagestyle{empty}
\newpage


\tableofcontents


\section{Introduction}

T-branes are among the most intriguing objects in String Theory. They were first realized as configurations of stacks of ordinary D-branes on which the vacuum expectation values of two of the worldvolume scalars and of the worldvolume flux are mutually non-commuting  \cite{Cecotti:2010bp} (see also \cite{Donagi:2003hh}). This non-Abelian piece of brane physics implies that geometric data alone is not enough to characterize them, which makes their behavior somewhat exotic from a model-building viewpoint. This is one of the main reasons for their intense investigation in the past few years \cite{Hayashi:2009bt,Chiou:2011js,Donagi:2011jy,Donagi:2011dv,Font:2013ida,Anderson:2013rka,DelZotto:2014hpa,Collinucci:2014qfa,Collinucci:2014taa,Marchesano:2015dfa,Cicoli:2015ylx,Carta:2015eoh,Heckman:2016ssk,Collinucci:2016hpz,Bena:2016oqr,Marchesano:2016cqg,Mekareeya:2016yal,Ashfaque:2017iog,Anderson:2017rpr,Bena:2017jhm,Collinucci:2017bwv,Cicoli:2017shd,Marchesano:2017kke,Anderson:2017zfm,Apruzzi:2018oge,Cvetic:2018xaq,Heckman:2018pqx,Apruzzi:2018xkw,Carta:2018qke,Marchesano:2019azf}.

In previous work \cite{Bena:2016oqr}, three of the authors and Minasian have shown that certain T-brane solutions preserving eight supercharges have a new description in the regime of parameters where the non-Abelian fields have large expectation values in string units, in terms of a collection of ordinary branes with commuting worldvolume scalars. This suggests a type of brane-brane duality, similar in spirit to the Myers effect \cite{Myers:1999ps,Constable:1999ac}, by which non-Abelian brane configurations in one regime of parameters correspond to charges and dipole moments of higher-dimensional Abelian branes in another regime of parameters. The crucial difference is that T-branes do not polarize, so the data of the non-Abelian description is not encoded in the dipole moments of some higher-dimensional brane, but rather in the curved shape of the holomorphic manifold wrapped by Abelian branes of the same dimension.

The evidence for the brane-brane duality conjectured in \cite{Bena:2016oqr} came via a rather indirect chain of reasoning that involved T-duality and the Myers effect, and depended crucially on certain details of the starting non-Abelian solutions. It is the purpose of this paper to show that this duality is a universal feature of all T-brane solutions, and to establish a way to systematically obtain the Abelian description of T-branes starting from their traditional non-Abelian description.

One highly non-trivial check of this duality is to match both brane actions, the Dirac-Born-Infeld (DBI) and the Wess-Zumino (WZ) ones, on the Abelian and the non-Abelian side. This guarantees that the energy density and all the induced brane charges agree in the two descriptions. Given the complexity of the non-Abelian brane action, at first glance such a match appears rather unlikely, except perhaps for very peculiar solutions. However, we will show that, upon taking into account curvature corrections to the Hitchin system and doing some rather non-trivial manipulations, one can establish this match systematically and in full generality. In particular, we will show that achieving the match is impossible unless the Abelian branes branch out into several components, according to a pattern dictated by the matrix structure of the T-brane worldvolume scalars.

This check, while non-trivial, is however not enough to establish the duality we are exploring here as a duality of string theory. To do so would include a check on the dynamics, such as the match of the low-energy fluctuations (zero-mode spectra) on both sides of the duality. This is something we do not study here, and it should be clear that when we refer to a duality, it is only established at the level of the kinematics of the two systems.\footnote{The duality of \cite{Bena:2016oqr} should however be considered a duality of string theory as it relies on T-duality, which is already a duality of string theory, and the Myers effect, for which the match of low-energy fluctuations have already been considered in \cite{Constable:1999ac}.}

To translate the T-brane non-Abelian data to the Abelian side, the strategy we will use is to first considerably simplify the expressions of the brane actions by making use of the supersymmetry equations, and then to formulate the duality match as a set of differential equations whose solution determines the holomorphic manifold wrapped by the branes in the Abelian description.

We will be able to establish the duality not only for T-branes preserving eight supercharges but also for T-branes preserving only four. The equations describing them are much more complicated (and also have much more interesting physics) and, to match all of their couplings, one needs to also take into account the next-to-leading order $\alpha'$ corrections to the supersymmetry equations and to the action.

We expect that effects coming from the finite size and nontrivial curvature of the cycles wrapped by T-branes will be suppressed as powers of ${\ell_s \over L_{CY}}$, where $L_{CY}$ is a typical size of the compactification manifold. Hence we will derive our dictionary restricting our analysis to T-branes extended on flat hyperplanes in a non-compact space. A common feature of the Abelian branes dual to these T-branes is that their worldvolume has no electromagnetic fluxes but has a non-trivial curvature. It is this curvature that encodes the information contained in the non-Abelian electromagnetic flux of the T-branes.

Even though one can construct T-branes with any kind of D$p$-branes\footnote{with $2\leq p \leq 7$ for 8 supercharges and $4 \leq p \leq 7$ for 4 supercharges.}, for traditional reasons we focus on T-branes constructed using D7-branes. We take them extended on $\mathbb{R}^{1,3}$ times a four-dimensional ``internal'' hyperplane. The standard formulation of their supersymmetric dynamics is in terms of a Hitchin-like system \cite{Hitchin:1986vp}, depending on a complex worldvolume scalar $\Phi$ (the so-called Higgs field) valued in the Lie Algebra of the gauge group, parameterizing the two transverse directions of the D7-branes.\footnote{We will be working in non-compact space and regard $\Phi$ as a scalar rather than as a top holomorphic form.} Calling the non-Abelian gauge field living on the stack of D7-branes and its field strength $A$ and $F_2$ respectively, minimal supersymmetry on $\mathbb{R}^{1,3}$ is preserved if and only if the following equations are satisfied:\footnote{Here we write the equations at leading order in $\alpha'$. We will discuss perturbative $\alpha'$ corrections in Section \ref{sec:4susies}.}
\begin{equation}
\label{eq:8hitchin}
\begin{split}
\bar \partial_{A} \Phi = &\, 0\, ,\\
F_2^{(0,2)} =&\, 0\, ,\\
2J \wedge F_2 + \star [\Phi, \Phi^\dagger] =&\,0\,,
\end{split}
\end{equation}
where we have fixed the complex structure on the brane worldvolume and have denoted the anti-holomorphic covariant derivative by $\bar \partial_{A}$, and the superscript on the two-form field strength selects its $(0,2)$ component. Moreover $J$ is the K\"ahler $(1,1)$-form of the internal part of the brane worldvolume, $\star$ is the Hodge-star operator on it, and $^\dagger$ denotes complex conjugation and matrix transposition. T-branes are generally defined as solutions of the above equations for $(A,\Phi)$ such that $[\Phi, \Phi^\dagger]\neq 0$. Because of the last equation in \eqref{eq:8hitchin}, T-branes necessarily involve non-primitive worldvolume fluxes, which are therefore not purely anti-self-dual. While solutions preserving only four supercharges have the pair $(A,\Phi)$ vary over a four-dimensional submanifold of the D7-brane worldvolume, for those preserving eight supercharges we only allow non-trivial profiles on a two-dimensional submanifold. Hence, for eight supercharges, the middle equation in \eqref{eq:8hitchin} is trivially satisfied.

We will outline a general methodology that, given a T-brane solution $(A,\Phi)$ of the Hitchin system \eqref{eq:8hitchin} as an input, allows us to construct the profiles of the Abelian branes dual to it. Such profiles will emerge as solutions to the differential equations arising from matching the brane actions, subjected to boundary conditions imposed by the given Lie Algebra structure of $\Phi$. We will find that such profiles describe holomorphic submanifolds of the target space, compatibly with supersymmetry (complex curves and complex surfaces for eight and four preserved supercharges respectively). However, since the duality involves non-holomorphic data through the last equation of \eqref{eq:8hitchin}, the complex structures of the brane worldvolumes that make supersymmetry manifest on both sides simultaneously are in general not equivalent.


Before beginning, it is important to understand the range of validity of the Abelian description of T-branes we are proposing and its possible overlap with the non-Abelian description above. Clearly, when the curvature of the hypersurface the Abelian branes wrap is large, the Abelian picture breaks down. At this scale, the non-Abelian picture in terms of the Hitchin system takes over, since the expectation values of the non-Abelian fields drop below the string scale.

Conversely, the non-Abelian description stops being valid for large values of flux densities (or equivalently of worldvolume-scalar commutators), because the non-Abelian DBI action, based on the symmetrized-trace prescription \cite{Tseytlin:1997csa,Myers:1999ps}, is known to disagree with string perturbation theory starting at order $({F}_2)^6$ \cite{Hashimoto:1997gm,Bain:1999hu}. This happens because for non-Abelian fields, certain terms containing derivatives of the fluxes can be rewritten as powers of the fluxes; this in turn makes the approximation of neglecting gradients of the flux while keeping powers of it, on which the DBI action is based, ill-defined \cite{Myers:2003bw}.

However, if we use minimal supersymmetry in four dimensions, and consider the on-shell DBI action, the validity of the non-Abelian description appears to extend to large expectation values of the non-Abelian fields. As we will discuss in Section \ref{sec:4susies}, this happens essentially because $\alpha'$ corrections to the last equation of \eqref{eq:8hitchin}, which are generally out of control because of the absence of holomorphicity, are related to corrections of the WZ action, and thus take a particularly well-controlled form. This, in turn, makes it possible to remove any ambiguity from the non-Abelian DBI action by writing it on-shell. We therefore argue that the symmetrized-trace prescription for the non-Abelian DBI action, when the latter is evaluated on a solution of the ($\alpha'$-corrected) Hitchin system, describes the physics correctly even for large non-Abelian-field vevs.

The rest of the paper is organized into two main Sections, \ref{sec:8susies} and \ref{sec:4susies}, which discuss T-branes solutions preserving eight and four supercharges respectively. Their structure is similar: We first describe the non-Abelian side and write down the corresponding DBI and WZ actions; then we do the same on the Abelian side; finally we derive the full set of conditions needed to match the actions on both sides. In Section \ref{ssec:8procedure} we propose an algorithmic procedure to use these duality conditions and to construct the Abelian dual of a given T-brane. In Section \ref{ssec:8example} we apply this method to revisit the example presented in \cite{Bena:2016oqr}. Finally, in Section \ref{sec:Discussion} we draw our conclusions and list a number of interesting questions left unanswered.

Since this paper involves several technically heavy computations, we report only a limited amount of details and results here, and accompany the paper with a Mathematica notebook \cite{MAT} which the reader is referred to for all the details.


\section{Eight supercharges} \label{sec:8susies}


In this section we will focus on generic T-branes preserving eight supercharges, and work out their dual Abelian description. In Section \ref{ssec:8duality} we will establish the general set of rules that this description needs to meet and discuss its features. We will then discuss in Section \ref{ssec:8procedure} how to find in principle explicit solutions for the dual picture, and finally in Section \ref{ssec:8example} we will apply the duality to a specific six-dimensional T-brane configuration, revisiting the example presented in \cite{Bena:2016oqr}.

\subsection{Duality} \label{ssec:8duality}

As we outlined in the Introduction, to establish our duality we have to demand that the energy density and the charge densities on the two sides are the same. These are complicated functions of the worldvolume coordinates, given by the DBI action and the couplings to the various Ramond-Ramond potentials in the WZ action. We start by deriving these expressions on both sides, and then proceed to lay down the general dictionary to match them.


\subsubsection{The non-Abelian side: Hitchin system}

Let us consider flat ten-dimensional space-time in Type IIB string theory and a flat stack of $N$ D7-branes. For T-brane configurations preserving eight supercharges, we can just restrict our attention to a $\mathbb{C}^2$ factor of the ten-dimensional space-time, which we parametrize by $w = w^1 + i w^2$ and $z = z^1 + i z^2$, and disregard the remaining six directions. The non-Abelian complex worldvolume scalar is written in terms of real scalars as $\Phi =\tfrac12(\phi^8 + i \phi^9)$, and the brane stack is embedded through the trivial holomorphic equation $f^{(\cancel{A})}(w,z) := z = 0$.\footnote{The superscripts $^{(\cancel{A})}$ and $^{(A)}$ denote quantities on the non-Abelian side and on the Abelian side respectively.}
We denote the only two real worldvolume coordinates we care about by $\sigma, s$, and choose to work in static gauge, meaning $\sigma=w^1$, $s=w^2$. Given the space-time complex structure chosen, this specifies the complex structure on the worldvolume such that $\sigma+is$ is the holomorphic coordinate.\footnote{Even though these appear as trivial remarks at this point, they are important for our later analysis, because the duality generically modifies the complex structure of the brane.} Having made this choice, we can write the equations governing the supersymmetric dynamics of this configuration in the form of a Hitchin system \eqref{eq:8hitchin}. As it will be more useful for us, let us express it in terms of real coordinates:
\begin{equation}\label{eq:hitchin8real}
  \begin{split}
    &D_{\sigma} \phi^8 = D_{s} \phi^9\,,\\
    &D_{s} \phi^8 + D_{\sigma} \phi^9 = 0\,,\\
    &F_{\sigma s} -i [\phi^8,\phi^9] = 0\,,
  \end{split}
\end{equation}
where we have used the fact that $A_{\bar{w}}=\tfrac12(A_\sigma+iA_s)$, $F_{\sigma s}=i(\partial_\sigma A_s-\partial_s A_\sigma)$, and we have defined $D_{\sigma(s)} := \partial_{\sigma(s)}+[A_{\sigma(s)},\cdot]$. In the following, we will be using the short-hand notation $\phi^c := - i[\phi^8, \phi^9]$.
As stated in the introduction, we focus on configurations where $\Tr \,\phi^8=\Tr\,\phi^9=0$, and where $\phi^c$ is along the Cartan of the gauge group.

The non-Abelian WZ action for a D$p$-brane, including Myers's terms \cite{Myers:1999ps} and curvature corrections \cite{Green:1996dd,Cheung:1997az,Minasian:1997mm} is:
\begin{equation}\label{eq:WZagen}
  S_{\WZ} = \mu_{p} \int \text{STr} \left[ \text{P} \left[e^{ i \lambda \iota_{\pmb{\phi}} \iota_{\pmb{\phi}}} \mathbf C \wedge e^B \right]\wedge e^{\lambda F_2}\right] \w \sqrt{\frac{\hat{\mathcal{A}}(2\pi \lambda R_T)}{\hat{\mathcal{A}}(2\pi \lambda R_N)}}\,,
\end{equation}
where  $\lambda = 2\pi \ell_s^2 = 2\pi \alpha'$ and $\mu_p=2\pi/(2\pi \ell_s)^{p+1}$, P denotes pull-back to the brane worldvolume, $\iota_{\pmb{\phi}}$ corresponds to contraction with the normal vector $\pmb{\phi}:=\{\phi^{p+1},\ldots,\phi^9\}$, $\mathbf{C}$ is the Ramond-Ramond polyform, $B$ is the NS-NS two-form, and STr indicates symmetric trace. The last factor corresponds to the curvature corrections, expressed in terms of the so-called $\hat{\mathcal{A}}$-roof genus, a polyform whose first non-trivial term is a four-form. For the moment, we can ignore these curvature terms, because only two dimensions of the brane worldvolumes we consider throughout this section are going to carry curvature. We will come back to them in Section \ref{ssec:4A}.

Let us choose for definiteness $p=7$. There are three types of RR fields that will have non-trivial couplings on both sides of  this duality:\footnote{We use the following notation here: $C_8 = \frac{1}{2!}\sum_{i,j} C_8^{(X^i X^j)}\w \d X^i \w \d X^j$, where $X^i=\{w^1,w^2,z^1,z^2\}$ are collective coordinates for the $\mathbb{C}^2$ subspace of the space-time. Note that, in our notation, $C_8^{(X^i X^j)}$ are \emph{six forms}.}
\begin{itemize}
  \item $C^{(w^1 w^2)}_8$. This is the $C_8$ component along $w^1$ and $w^2$, arising from the trivial pull-back of the static gauge $\sigma=w^1$ and $s=w^2$.

 \item $C^{(z^1 z^2)}_8$. These components of $C_8$ arise in two ways: Non-Abelian pull-back which uses $\lambda\Phi$ as transverse coordinate, and non-Abelian contraction $\iota_{\pmb{\phi}} \iota_{\pmb{\phi}}$ followed by wedge with $F_2$.

  \item \emph{Mixed}. In principle there could be couplings to forms with mixed legs, of the type $C^{(w^1 z^2)}$, in which we have a non-trivial, non-Abelian pull-back corresponding to the coordinate $z^2$, and a trivial one corresponding to the coordinate $w^1$. These terms will contain one power of the scalar $\phi^9$. Since this field is traceless all these couplings are zero.
\end{itemize}

The final expression for the WZ action on the non-Abelian side can be written as (see \cite{MAT} for details)
\begin{equation}\label{eq:8naWZ}
  S^{(\cancel{A})}_{\WZ} = \mu_7 \int \left( N C_8^{(w^1 w^2)} + \lambda^2 C_8^{(z^1 z^2)} \Tr\left\{(D_{\sigma} \phi^8)^2+(D_{s}\phi^8)^2 -\phi^c F_{\sigma s}\right\} \right) \wedge \d \sigma \wedge \d s\,,
\end{equation}
where we have used the first two equations of \eqref{eq:hitchin8real} to get rid of expressions containing $\phi^9$. Note that the term $P[C_6] \w \Tr\{F_2\}$, contained in \eqref{eq:WZagen}, is zero for a T-brane solution because $\Tr\,\{F_2\} = \Tr\,\{[\Phi,\Phi^\dagger]\} = 0$.

The non-Abelian DBI  action \cite{Myers:1999ps} together with the curvature corrections \cite{Bachas:1999um}, has the form\footnote{We use $\xi$ as a collective symbol for worldvolume coordinates.}
\begin{equation}\label{eq:DBI}
  \begin{split}
    S_{\DBI} &= - \mu_p \int \d^{p+1} \xi e^{-\phi} \STr \sqrt{-\det(P[E_{\alpha\beta} + E_{\alpha a}(Q^{-1} - \delta)^{ab}E_{b\beta}] + \lambda F_{\alpha \beta}) \det Q}\\
    &\qquad \qquad \times \Big( 1 - \frac{1}{24} \frac{(2\pi \lambda)^2}{32\pi^2} \big( (R_T)_{\alpha\beta\gamma\delta}(R_T)^{\alpha\beta\gamma\delta} - 2(R_T)_{\alpha\beta}(R_T)^{\alpha\beta} \\
    &\qquad\qquad\qquad\qquad\qquad\qquad- (R_N)_{\alpha\beta ab}(R_N)^{\alpha \beta ab} + 2 \bar{R}_{ab} \bar{R}^{ab} + \mathcal{O}(\lambda^4)\big)\Big)\,,
  \end{split}
\end{equation}
where $\phi$ is the dilaton and
\begin{equation}
  \begin{split}
    Q^a_{\ b} = \delta^a_{\ b} + i \lambda [\phi^a, \phi^c] E_{cb}\,,\quad E_{\mu \nu} = G_{\mu \nu} + B_{\mu \nu}\,,
  \end{split}
\end{equation}
with $G_{\mu\nu}$ the space-time metric.

In the above, the indices $\mu,\nu$ are space-time indices, running from $0$ to $9$, $\alpha,\beta$ are worldvolume indices, running from $0$ to $p$, and $a,b$ are transverse indices, running from $p+1$ to $9$. The worldvolume Riemann and Ricci tensors are $(R_T)_{\alpha\beta\gamma\delta}$ and $(R_T)_{\alpha\beta}$ respectively, and  $R_N$ is the curvature two-form of the $SO(9-p)$ normal bundle. As before, these curvature terms vanish throughout this section. We will discuss them in more detail in Section \ref{ssec:4A}, where they are going to play a key r\^ole.

Using the embedding of the D7-brane configuration outlined above, one can show that the DBI  action \eqref{eq:DBI} takes the form (see \cite{MAT} for details)
\begin{equation}\label{eq:8nADBI}
  S^{(\cancel{A})}_{\DBI} = - \mu_7 \int \d^{6} \xi \, \d \sigma\, \d s\, e^{-\phi} \,\left[ N + \lambda^2 \Tr\{(D_{\sigma}\phi^8)^2 +(D_{s}\phi^8)^2-\phi^cF_{\sigma s} \} \right]\,,
\end{equation}
where the first two equations of \eqref{eq:hitchin8real} have again been used to simplify this expression.


\subsubsection{The Abelian side: Curved branes}

With \cite{Bena:2016oqr} as a guide, we look for a dual configuration involving ordinary D-branes wrapping curved, non-trivially embedded Riemann surfaces, without a worldvolume gauge field. The space-time is still parametrized by complex coordinates $w$ and $z$, and the brane worldvolume by real coordinates $\sigma$ and $s$. If we insist on keeping supersymmetry manifest in our dual description, the brane embedding should be described as the zero-locus of a holomorphic function $f^{(A)}(w,z)$. Away from possible poles, we may always invert the latter equation and get a relation $z=z(w)$, which is better suited for the computations below. Since we fixed the real worldvolume coordinates on both sides of the duality to be $\sigma,s$, we also need to specify the relation $(\sigma,s)\leftrightarrow(w^1,w^2)$, which we chose to be the identity on the non-Abelian side. This relation, in turn, will determine a choice of complex structure on the dual brane, which will in general not be holomorphically related to the one on the non-Abelian stack. This is to be expected, because the duality involves non-holomorphic data through the last equation of \eqref{eq:hitchin8real}.\footnote{In this construction method, the freedom of reparametrization of the brane-worldvolume coordinates shows up as an overall modification in the relationship $(\sigma,s)\leftrightarrow(w^1,w^2)$ on both sides of the duality. Of course, to keep supersymmetry manifest, such modification needs to respect the holomorphic structure.}

To establish the duality we will now derive the WZ action and the DBI action on this Abelian side, and compare them to the corresponding expressions \eqref{eq:8naWZ} and \eqref{eq:8nADBI} of the non-Abelian side. Let us start with the WZ action. Using the general formula \eqref{eq:WZagen}, the fact that we have no worldvolume flux, and the holomorphicity of the embedding (i.e. $\partial_{\bar w}z=0$), we find (see \cite{MAT} for details):
\begin{equation}\label{eq:8aWZ}
  S^{(A)}_{\WZ} = \mu_7 \int \det \mathcal{J} \left[C_8^{(w^1 w^2)} + C_8^{(z^1 z^2)} |\p_w z|^2 + (\p_w z)\, C_8^{(z \bar w)} + (\p_{\bar w} \bar z)\, C_8^{(\bar z w)}\right] \wedge \d \sigma \wedge \d s\,,
\end{equation}
where $\mathcal{J}$ is the Jacobian matrix that pulls the space-time coordinates $X^i = \{w^1, w^2\}$ back to the worldvolume coordinates $\xi^\alpha =\{\sigma, s\}$: $\mathcal{J} = [\partial X^i/\partial \xi^\alpha]$.

We can see that there is a discrepancy between the Abelian couplings \eqref{eq:8aWZ} and the non-Abelian ones \eqref{eq:8naWZ}, coming from the sources for the mixed forms  $C_8^{(z \bar w)}$ and $C_8^{(\bar z  w)}$. Furthermore, one cannot impose these sources to be zero, since this would prevent the $C_8^{(z \bar z)}$ coupling to match.
We also know that charge conservation imposes that the Abelian branes dual to a T-brane realized on $N$ D7 branes must also have charge $N$.\footnote{One may attempt to formulate the dual description in terms of a single Abelian brane wrapped $N$ times, as done in \cite{Bena:2016oqr}. This corresponds to fixing det$\mathcal{J}=N$, which still would not be sufficient to match the mixed terms.}

Both of these issues can be resolved once we realize that the dual Abelian brane configuration is made of $N$ different strands, with WZ actions given by \eqref{eq:8aWZ}, each of them separated by a phase $z \to z e^{i\chi_k}$, for the $k^{\text{th}}$ strand. The action to be matched with the non-Abelian side is then the sum over the $N$ individual strands. The mixed terms get cancelled under this sum by appropriate choices of phase, for instance $\chi_k = 2 \pi k/N$.

Explicitly this procedure means that the correct expression to match with \eqref{eq:8naWZ} is instead
\begin{equation}\label{eq:finalSAWZ}
  \begin{split}
    S^{(A)}_{\WZ} &= \sum_{k=0}^{N-1} \mu_7 \int \det \mathcal{J} \left[C_8^{(w^1 w^2)} + C_8^{(z^1 z^2)} |\p_w z|^2 \right.\\
    & \qquad \qquad \qquad \qquad \qquad \qquad \left. + e^{\frac{2 \pi i k}{N}} \p_w z\, C_8^{(z \bar w)} + e^{-\frac{2 \pi i k}{N}} \p_{\bar w} \bar z\, C_8^{(\bar z w)}\right] \wedge \d \sigma \wedge \d s\\
    &=N \mu_7 \int \det \mathcal{J}\left[C_8^{(w^1 w^2)} + C_8^{(z^1 z^2)} |\p_w z|^2 \right] \wedge \d \sigma \wedge \d s\,.
  \end{split}
\end{equation}
We will discuss the interpretation of this multi-branched Abelian brane when we revisit the explicit example of \cite{Bena:2016oqr}, in Section \ref{ssec:8example}.

Let us now turn to the DBI  action, which, using the general formula \eqref{eq:DBI}, can be written as (see \cite{MAT} for details):
\begin{equation}\label{eq:8ADBI}
  \begin{split}
    S^{(A)}_{\DBI} &= - \sum_{k=0}^{N-1} \mu_7 \int \d^{6} \xi \d \sigma \d s e^{-\phi}\,  |\det \mathcal{J}| \left(1+|\p_w z|^2\right)\\
    &= - \mu_7 N \int \d^{6} \xi \d \sigma \d s e^{-\phi}\,  |\det \mathcal{J}| \left(1+|\p_w z|^2\right)\,,
  \end{split}
\end{equation}
where again we have used the holomorphicity of the embedding and the absence of flux. Notice that, the DBI  action does not come with mixed terms before the sum over each strand of the Abelian brane. Also, note that, even though the worldvolume of the Abelian brane is curved, the curvature terms in the DBI \eqref{eq:DBI} are still vanishing. This is because the curved part of the worldvolume is a Riemann surface, and thus the various curvature quantities appearing in \eqref{eq:DBI} have the simple form
\begin{align}
(R_T)_{\alpha \bar \beta \gamma \bar \delta} (R_T)^{\alpha \bar \beta \gamma \bar \delta} &= \left(\Omega{\overline \Omega}\right)^2 h^{-4}\,,\\
(R_N)_{a \bar b \gamma \bar \delta} (R_N)^{a \bar b \gamma \bar \delta} &=\left(\Omega{\overline \Omega}\right)^2 h^{-4}\,,\\
(R_T)_{\alpha \bar \beta} (R_T)^{\alpha \bar \beta} &=-\left(\Omega{\overline \Omega}\right)^2 h^{-4}\,,\\
\bar R_{a \bar b} \bar R^{a \bar b} &= -\left(\Omega{\overline \Omega}\right)^2 h^{-4}\,,
\end{align}
where $\Omega$ is the only surviving component of the second fundamental form, which we will discuss in more detail in Section \ref{ssec:4A}, and $h$ is the hermitian metric on the Riemann surface.


\subsubsection{Duality dictionary}\label{sssec:dd8}

We are now in the position to compare the two pairs of actions: The non-Abelian WZ \eqref{eq:8naWZ} with the Abelian WZ \eqref{eq:finalSAWZ}, and similarly the DBI  actions \eqref{eq:8nADBI} and \eqref{eq:8ADBI}. Since we are going to establish a brane-brane duality, we must do it independently of closed-string data. Hence we have to match each $C_8$-component in the WZ actions. This brings us to the following two real conditions
\begin{align}
    &\det \mathcal{J} = 1\,, \label{eq:8duality1}\\
    & |\p_w z|^2 = \frac{\lambda^2}{N} \Tr\{(D_{\sigma}\phi^8)^2 +(D_{s}\phi^8)^2- \phi^c F_{\sigma s} \} \,, \label{eq:8duality2}
\end{align}
which also imply equality of the DBI  actions.

In order to get acquainted with these conditions, let us start by understanding what they mean for ordinary D-branes with $\phi^c=0$. While Equation \eqref{eq:8duality1} can be trivially solved by choosing a static gauge, Equation \eqref{eq:8duality2} can be rewritten as
\begin{equation}\label{eq:8duality2nonT}
|\p_w z|^2 = \frac{4\lambda^2}{N} \Tr|\partial_{w}\Phi|^2 \,,
\end{equation}
where we have simultaneously diagonalized $\phi^8$ and $\phi^9$. This equation simply follows from the equivalence of two different descriptions of an Abelian system of $N$ D7-branes: On one hand we can describe it as a stack of $N$ D7-branes placed at $z=0$ with a non-trivial vev for the Higgs field along the Cartan of the gauge group, and this corresponds to the r.h.s.~of \eqref{eq:8duality2nonT}; on the other hand, the l.h.s.~of \eqref{eq:8duality2nonT} corresponds to describing the same system by the embedding function of each of the $N$ components (strands) of the stack. This equivalence is expressed by the identification
\begin{equation}\label{eq:8duality2nonTbis}
z_k(w) \longleftrightarrow \frac{2\lambda}{\sqrt{N}}\Phi_{kk}(w) \,,
\end{equation}
where $k=1,\ldots,N$ labels the strands, and where the r\^ole of the eigenvalues of $\Phi$ as transverse-deformation fields is apparent.

In Equation \eqref{eq:8duality2} we have the trace of a $N\times N$ matrix, but the $N$ dependence of such a trace highly depends on the details of the T-brane configuration we focus on. A particular configuration of T-brane type where we can explicitly extract the $N$ dependence of the trace is one in which $\phi^8,\phi^9,\phi^c$ are proportional to the generators $\{\Sigma_a\}_{a=1,2,3}$ of an $\mathfrak{su}(2)$ subalgebra of the gauge algebra. For the $\mathfrak{su}(N)$ gauge algebra, this happens for example when $\Phi\propto\sum_i\sqrt{i(N-i)}E_{\alpha_i}$, where the sum extends over the set of simple roots $\{\alpha_i\}_{i=1,\ldots,N-1}$, with associated generators $E_{\alpha_i}$. This corresponds to the principal nilpotent embedding, where the $N$ dependence of the trace is given by: $\Tr(\Sigma_a\Sigma_b)=\delta_{ab}N(N^2-1)/3$.\footnote{The explicit example we will discuss in Section \ref{ssec:8example} belongs to this class of T-brane configurations.}

There is another important point we want to emphasize here. The non-Abelian side has a manifestly supersymmetric description in terms of the Hitchin system (\ref{eq:8hitchin}), where we fixed the complex structure on the brane stack by choosing $w =F^{(\cancel{A})}(\sigma,s) = \sigma + i s$. Similarly, on the Abelian side, supersymmetry has been made manifest by working with an holomorphic function $z=z(w)$. Solving equations \eqref{eq:8duality1} and \eqref{eq:8duality2} in particular amounts to specifying a new relation of the type $w=F^{(A)}(\sigma,s)$. The trivial relation $F^{(A)}(\sigma,s)=\sigma+is$, chosen for the Hitchin system, would still solve Equation \eqref{eq:8duality1}, that can also be written as
\begin{equation}\label{eq:8duality11}
\det \mathcal{J}=- \Im [\partial_\sigma w \partial_s \bar{w}]=1\,.
\end{equation}
However, as we will see explicitly later in an example, this relation would in general be incompatible with Equation \eqref{eq:8duality2}, thereby forcing us to choose a different  $F^{(A)}(\sigma,s)$ solving \eqref{eq:8duality1}. Such a relation will in general alter the complex structure chosen on the brane in the non-Abelian picture, in the sense that $F^{(\cancel{A})}(\sigma,s)$ and $F^{(A)}(\sigma,s)$ are linked by a non-holomorphic transformation:
\begin{equation}
F^{(A)} = F^{(A)} \left(F^{(\cancel{A})},\bar{F}^{(\cancel{A})}\right)\,.
\end{equation}

We would also like to make an observation regarding the mixed terms in Equation \eqref{eq:8aWZ} and their cancellation. As we mentioned in the previous subsection, these mixed terms are eliminated by a relative rotation $z \to z e^{2\pi i k/N}$ for the $k^{\rm th}$ of the $N$ strands of this Abelian brane. This is possible because the profile of each of these strands has to satisfy Equation \eqref{eq:8duality2}, which is invariant under a constant phase shift of $z$.

In the next subsection, we will prescribe the steps one would have to follow in order to derive, given an explicit T-brane solution on the non-Abelian side, the embedding of the dual Abelian brane satisfying equation \eqref{eq:8duality2}.


\subsection{A solving procedure} \label{ssec:8procedure}

The discussion of the previous sections suggests the existence of a brane-brane duality, governed by the two real conditions written in \eqref{eq:8duality1} and \eqref{eq:8duality2}. We would now like to study them in more detail: In particular, we are going to outline a systematic procedure that, given the non-Abelian-brane description as an input of the problem, may be used for the derivation of the Abelian-brane dual of the same system.

The starting point is to solve Equation (\ref{eq:8duality1}), which means fixing the function $F^{(A)}(\sigma,s)$. As is clear from \eqref{eq:8duality11}, this equation does not have a unique solution. Consider for example the Ansatz
\begin{equation}\label{eq:wansatz}
  w = F^{(A)}(\sigma,s)= h(\sigma) e^{i s / (h^\prime(\sigma) h(\sigma))}\,,
\end{equation}
which solves Equation (\ref{eq:8duality1}) for any $h(\sigma)$. In the above, the symbol $^\prime$ indicates the derivative with respect to $\sigma$. However, the symmetry of Equation (\ref{eq:8duality11}) under the interchange of $w \leftrightarrow \bar{w}$ and $\sigma \leftrightarrow s$ makes it possible to generate yet another Ansatz that solves (\ref{eq:8duality1}). We have not attempted to study the solution space of \eqref{eq:8duality1}, in particular how large this is. Rather, we stick with Ansatz \eqref{eq:wansatz}, which will prove to be a viable choice in our working example, and proceed with the solving algorithm.

The second step would be to solve the differential equation (\ref{eq:8duality2}), whose unknown is the function $z(w)$. However, the r.h.s.~of (\ref{eq:8duality2}) is explicitly expressed in terms of $\sigma,s$, and hence we would need to invert \eqref{eq:wansatz} and write $\sigma,s$ in terms of $w,\bar w$, in order to be able to solve (\ref{eq:8duality2}) for $z(w)$. But this is impossible until we specify $F^{(A)}(\sigma,s)$ completely. Nevertheless, we can still proceed by making an important observation. Equation (\ref{eq:8duality2}) implies that the expression in its r.h.s.~is an absolute value of a holomorphic function of $F^{(A)}$, and not just of a holomorphic function of $F^{(\cancel{A})}$, as can be easily checked. Moreover, the logarithm of the absolute value of a holomorphic function is harmonic. We can now use the expression of the Laplace operator in terms of $w,\bar w$ to convert this harmonicity condition into a differential equation constraining the Ansatz for $F^{(A)}(\sigma,s)$, and fixing  the $w$-dependence on $\sigma,s$ completely.

Summarizing, the algorithm goes as follows:
\begin{enumerate}
  \item Start with an explicit solution to the Hitchin system, Eq.~(\ref{eq:hitchin8real}).
  \item Fix an Ansatz solving Eq.~(\ref{eq:8duality1}). \label{it:8d1}
  \item Rewrite the harmonicity condition
  \begin{equation}\label{eq:harmonicity}
    \partial_w \partial_{\bar{w}} \log \left(\Tr\{(D_{\sigma}\phi^8)^2 +(D_{s}\phi^8)^2-\phi^cF_{\sigma s}\}\right) = 0
  \end{equation}
 in terms of $F^{(A)}(\sigma,s)$, by using the expression $\partial_w = \frac{i}{2} (\partial_\sigma \bar{F}^{(A)} \partial_s - \partial_s \bar{F}^{(A)} \partial_\sigma)$, which follows from chain rule and Eq. \eqref{eq:8duality11}. Now solve \eqref{eq:harmonicity} for $F^{(A)}(\sigma,s)$, using the explicit solution to the non-Abelian system (\ref{eq:hitchin8real}).

 \item If no solution is found, repeat Item \ref{it:8d1} by finding a different Ansatz.

 \item Having fixed the form of $F^{(A)}(\sigma,s)$, we can now express $\sigma,s$ in terms of $w,\bar w$ using \eqref{eq:wansatz}, finally rewrite Eq.~(\ref{eq:8duality2}) as a differential equation for $z = z(w)$, and solve it.

\end{enumerate}

There is a little caveat here. Assuming that $z$ is a holomorphic function of $w$ is certainly not general enough, in the sense that poles may arise on the $w$-plane (as we will see in the next section). However, the embedding of the Abelian brane can be globally described by a holomorphic function $f^{(A)}(w,z)$. In this description, the relevant quantity $\partial_w z$, which appears in the actions \eqref{eq:finalSAWZ} and \eqref{eq:8ADBI}, should be replaced by $-\partial_wf^{(A)}/\partial_z f^{(A)}$. This can be seen as follows: On the Abelian-brane worldvolume, $f^{(A)}(w,z)=0$ and is constant. Therefore its total derivative with respect to any coordinate on the brane must vanish, implying by chain rule that $\partial_\xi w\partial_w f^{(A)}(w,z)+\partial_\xi z\partial_z f^{(A)}(w,z)=0$ for any complex worldvolume coordinate $\xi$. Going to static gauge, $\xi=w$, we obtain the desired relation.

Before ending this section, we would like to stress that we have not investigated whether the above algorithm leads us to a unique solution for the Abelian side, given a specific input solution to the Hitchin system. We would expect, though, that every solution for $z=z(w)$ of Equations \eqref{eq:8duality1} and \eqref{eq:8duality2} would be related to each another by holomorphic changes of the $w$-variable, thereby leaving the brane physics invariant. It would be interesting to verify this statement.


\subsection{Explicit example}\label{ssec:8example}

In this section we will revisit the example of \cite{Bena:2016oqr}, in which the Abelian dual of a nilpotent T-brane configuration was derived. We will show how the procedure outlined in the previous section can be used to deduce and generalize the final result of \cite{Bena:2016oqr}.

The non-Abelian T-brane solution considered in \cite{Bena:2016oqr} is given by
\begin{equation}\label{eq:nasol8}
  \phi^8 = g(\sigma) \Sigma_1 \,,\quad \phi^9 = g(\sigma) \Sigma_2\,,\quad A_{s} = i\frac{g'(\sigma)}{2g(\sigma)} \Sigma_3\,;\qquad g(\sigma) = \frac{C}{2 \sinh(C \sigma)}\,,
\end{equation}
where the $\Sigma_i$ form an $N$-dimensional representation of the $\mathfrak{su}(2)$ algebra and $C$ is a real constant with the dimension of a mass. This is the input to our problem.

The next step of our procedure is to select an Ansatz solving (\ref{eq:8duality1}), and we choose \eqref{eq:wansatz}. Now we must evaluate and solve (\ref{eq:harmonicity}), which, after some manipulations, amounts to finding $h(\sigma)$ that satisfies
\begin{equation}\label{eq:UglyEquation}
   \big(h'(\sigma)\big)^2 \Xi - h(\sigma) \big( h'(\sigma) \Upsilon + h''(\sigma) \Xi \big) = 0\,,
\end{equation}
with
\begin{equation}
  \begin{split}
    \Xi &= \sinh(C \sigma) \cosh(C \sigma) \left(2 + \cosh(2C\sigma)\right)\left(5 + \cosh(2C\sigma)\right)\,,\\
    \Upsilon &= C [9 + \left(2+\cosh(2C \sigma)\right)\left(4 \cosh(2C\sigma) - 1\right)]\,.
  \end{split}
\end{equation}
The differential equation \eqref{eq:UglyEquation} happens to have a fairly simple solution:
\begin{equation}\label{eq:hsol}
  h(\sigma) = \frac{3^{1/4} \sinh(C \sigma)}{C [2 + \cosh(2C\sigma)]^{1/4}}\,,
\end{equation}
where integration constants have been chosen such that $h(\sigma) \to \sigma$ as $C \to 0$. We will return to discuss the interpretation of these integration constants later on.

We now have $F^{(A)}(\sigma, s)$ explicitly given, and we are able to reformulate the r.h.s.~of (\ref{eq:8duality2}) in terms of $w$ and $\bar w$. For the particular solution at hand, the latter is independent of $s$. Moreover we have $w \bar w = h(\sigma)^2$. Therefore we can rewrite (\ref{eq:8duality2}) as:
\begin{equation}\label{eq:8duality21}
  |\partial_w z|^2 = \lambda^2 \frac{C^4 (N^2 - 1)}{4} \;\frac{1 + \frac{2}{3}\sinh^2(C \sigma)}{\sinh^4(C \sigma)}\; \tooo_{N \gg 1}\; \frac{\lambda^2 N^2}{4 h^4(\sigma)} = \frac{\lambda^2 N^2}{4 (w \bar w)^2} = \left|\frac{\lambda N}{2w^2}\right|^2\,,
\end{equation}
where, consistently with the procedure followed in \cite{Bena:2016oqr}, we took a large-$N$ approximation and hence have neglected $1/N^2$ corrections.

\begin{figure}[h!]
  \centering
  \begin{tabular}{cc}
    \includegraphics[scale=.5]{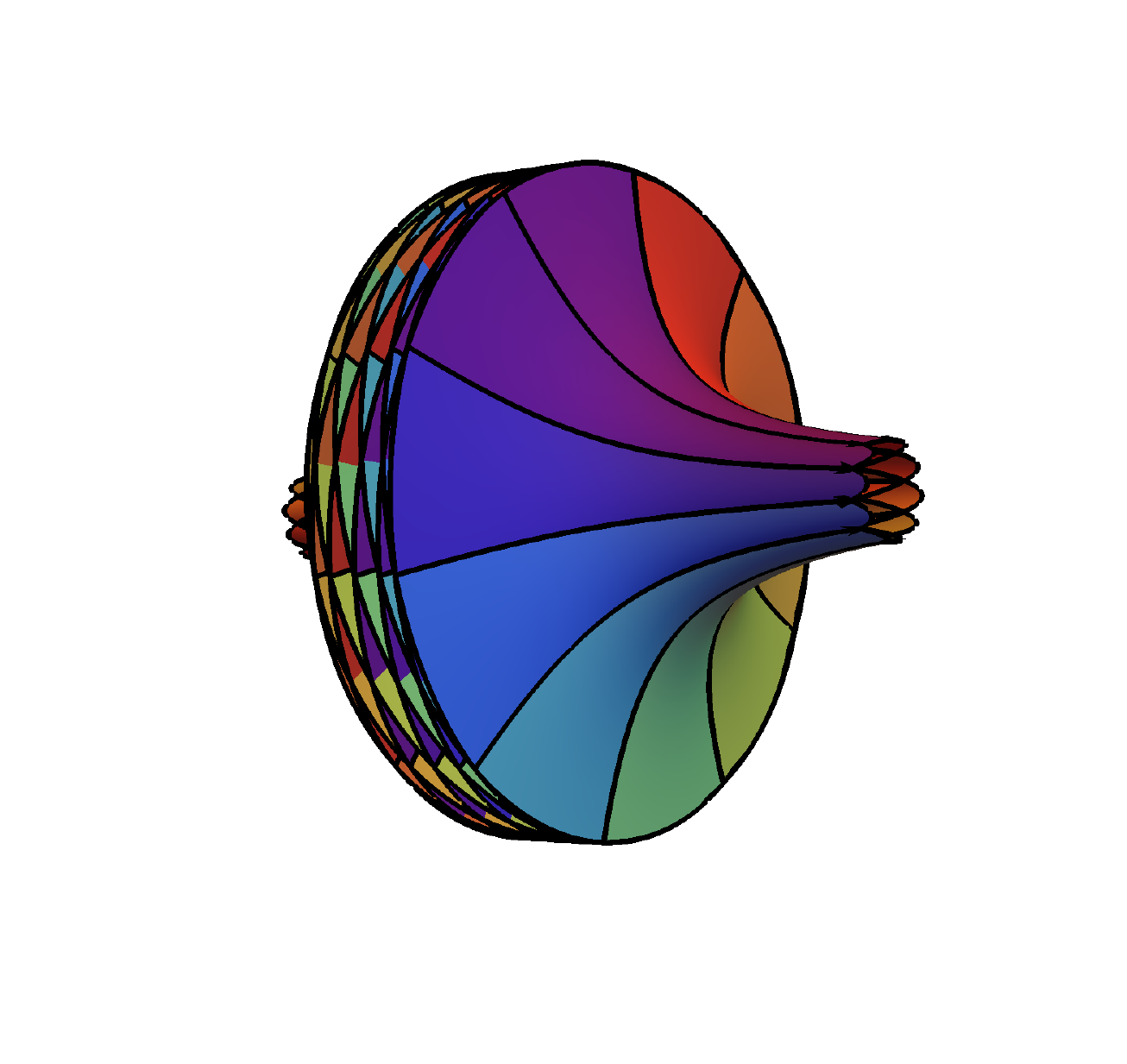}
    &
    \raisebox{0.16\height}{\includegraphics[trim={0 0 0 80},clip,scale=.5]{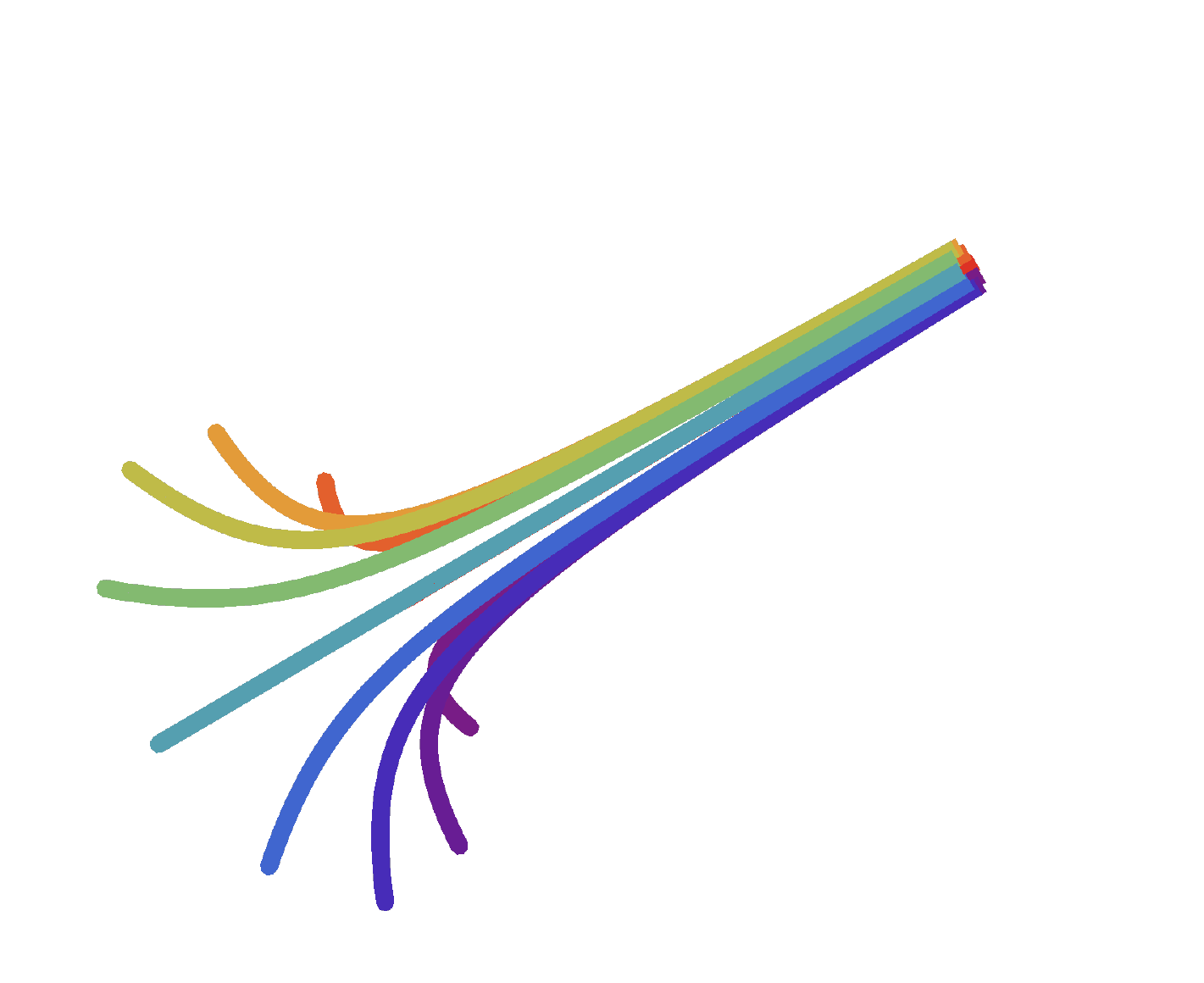}}
  \end{tabular}
  \caption{\it A plot of $N=10$ strands, projected from four dimensions down to $(w^1, w^2, z^2)$, with $C=0.1$. To the left the strands are plotted in their complete form, while on the right we have chosen to plot each strand along a single line only. In both pictures the strands separate into a disk in the $w$-plane. The cylindrical pole (which we have made more pronounced in the right picture) is the remainder of the $z$-plane after the projection down to three dimensions along its imaginary part. At the core of the pole, where all the strands merge, the non-Abelian description is reliable; as the strands open up, the Abelian picture takes over. Each of the strands are individually colored, and their misalignment is caused by the phase shift.}
  \label{fig:nsheets}
\end{figure}

We can now solve the differential equation for $z(w)$, putting the integration constant (which gives position of the-center-of-mass of the Abelian brane) to zero. Equation \eqref{eq:8duality21} leaves us with the freedom of a constant phase, which can be chosen differently for each strand of the solution. Calling this phase $\chi_k$ for the $k^\text{th}$ strand, we have
\begin{equation}\label{eq:embed}
  z = \frac{\lambda N}{2 w} e^{i\chi_k}\,.
\end{equation}
As described in Section \ref{ssec:8duality}, in order to cancel the mixed terms in the action, one convenient choice is $\chi_k = 2 \pi k/N$ for the $k^\text{th}$ strand.\footnote{In \cite{Bena:2016oqr}, these phases were mistakenly chosen all equal, thus leaving mixed terms of the Abelian WZ action uncanceled. Note also that, in the method of \cite{Bena:2016oqr}, in contrast to the present one, this freedom of phase arose as the choice of integration constant of a first-order differential equation.} In solving the differential equation (\ref{eq:UglyEquation}), we have fixed two positive, real integration constants, $C_1$ and $C_2$, in such a way as to recover a regular shape \eqref{eq:embed}, without branch cuts. Different choices of such constants would amount to the holomorphic change of variable $w \to C_1 w^{C_2}$, which of course carries no physical information. A three-dimensional projection of this shape has been drawn in Figure \ref{fig:nsheets}, plotted for ten strands.

We would like to end this section by a few remarks. In \cite{Bena:2016oqr} the family of solutions above (parametrized by $C$) was obtained using a chain of dualities: A first T-duality, followed by a brane-polarization effect, and a final T-duality. However, the last step was only taken for the $C=0$ representative of the family. The direct construction we obtain in this paper allows us to instead derive the Abelian dual of the whole family of solutions.

Also, note that the constant $C$ does not appear in the embedding expression \eqref{eq:embed}, but is hidden in the expression of $w=F^{(A)}(\sigma, s)$. In other words, $C$ does not affect the shape of the Abelian brane, which is thus the same for every solution in the family, but rather it determines the complex structure on its worldvolume, through Equation \eqref{eq:hsol}.

Another important improvement of our approach is that it allows us to derive an \emph{exact-in-$N$} Abelian dual. Indeed, nothing prevents us from solving Equation \eqref{eq:8duality21} without making any large-$N$ approximation, thereby obtaining the correct factor of $\sqrt{N^2-1}$ rather than $N$ in the shape \eqref{eq:embed}. While this seems perfectly consistent in the present context, it would not be so within the framework of \cite{Bena:2016oqr}, where the Abelian dual was derived using a T-dual Type-IIA brane configuration. The reason is that the T-dual D6-D8 system involved a compactly supported worldvolume flux on the funnel-shaped D8-brane worldvolume, which is integrally quantized and accounts for the $N$ units of D6-brane charge. It is precisely this quantization condition that forces the non-Abelian and the Abelian pictures to match only approximately at large $N$. Here, on the contrary, we have never assumed the existence of a T-dual frame for our D7-brane configurations, and we have derived the duality in a completely non-compact context directly in Type IIB string theory. Therefore, we do not have to impose any further consistency constraints from tadpoles or flux quantization conditions, and consequently there is no reason to expect our duality not to be exact in $N$. Extending our work to compact settings will certainly induce further requirements, and consequently the duality match may need a large-$N$ expansion to work.

Finally, one may be worried that the Abelian dual solution we just constructed is made of $N$ independent branes, thus giving rise to a $U(1)^N$ gauge symmetry, as opposed to the single $U(1)$ left unbroken by the corresponding non-Abelian solution. However, we can easily see that the different strands are all frozen by their boundary conditions, and all join at two points. Indeed, from Equation \eqref{eq:embed} it follows that the various strands never meet, except at infinity. To see what happens near infinity in either complex plane, it is convenient to replace $\mathbb C_z \times \mathbb C_w$ with $\mathbb{CP}^1_{[z_1:z_2]} \times \mathbb {CP}^1_{[w_1 : w_2]}$. Working in units of $\lambda$, in the homogeneous coordinates of the two $\mathbb {CP}^1$'s the equation for the surface \eqref{eq:embed} becomes
\begin{align}
z_1 w_1 = \frac{N}{2} e^{\frac{2 \pi i k}{N}} z_2 w_2\,.
\end{align}
So far we have been looking at the patch where $z_2 $ and $w_2$ are both non-zero and therefore it is possible to set them to 1. However, there are four affine patches in total. Calling the patch where the coordinate $x$ does not vanish $\mathcal U_{x}$, we have
\begin{align}
& \mathcal U_{z_2} \cap \mathcal U_{w_2}\ \rightarrow \ z w = \frac{N}{2} e^{\frac{2 \pi i k}{N}} \,,\\
&\mathcal U_{z_1} \cap \mathcal U_{w_2}\ \rightarrow \  w = \frac{N}{2} e^{\frac{2 \pi i k}{N}} \tilde z \,,\\
&\mathcal U_{z_2} \cap \mathcal U_{w_1}\ \rightarrow \ z  = \frac{N}{2} e^{\frac{2 \pi i k}{N}}\tilde w \,,\\
&\mathcal U_{z_1} \cap \mathcal U_{w_1}\ \rightarrow \ 1 = \frac{N}{2} e^{\frac{2 \pi i k}{N}} \tilde z \tilde w\,.
\end{align}
Here $z = z_1/z_2$, $\tilde z = z_2/z_1$ and similarly for $w$. We see that there is no intersection in $ \mathcal U_{z_2} \cap \mathcal U_{w_2}$ and in $ \mathcal U_{z_1} \cap \mathcal U_{w_1}$, but all strands join at $\tilde z = w = 0 $ in $ \mathcal U_{z_1} \cap \mathcal U_{w_2}$ and at $\tilde w = z = 0 $ in $ \mathcal U_{z_2} \cap \mathcal U_{w_2}$. This confirms that there is only one brane with different strands visible in the patch we were considering before.

This example suggests that, in general, the Abelian dual of any T-brane configuration would be specified by a meromorphic embedding expression $z=z(w)$, and by a strand structure where all the different strands would be frozen by boundary conditions.


\section{Four supercharges}\label{sec:4susies}

Let us now turn attention to T-brane configurations preserving only four supercharges. We will follow the same strategy as in the previous section to derive the general conditions that corresponding Abelian configurations need to meet. We first discuss in Section \ref{ssec:4nA} the non-Abelian side with the corresponding Hitchin system of equations, deriving the general expressions of WZ and DBI  actions, and distinguishing the various types of induced (lower-dimensional) brane charges. We then turn to the analysis of the Abelian side in Section \ref{ssec:4A}, where we follow a similar structure in the derivation of the general WZ and DBI  actions. Finally, in Section \ref{ssec:dual4}, we provide the duality dictionary relating the two brane systems.


\subsection{The non-Abelian side}\label{ssec:4nA}

This time we retain a $\mathbb{C}^3$ factor of the flat ten-dimensional space-time of Type IIB string theory, parametrized by $v = v^1 + i v^2$, $w = w^1 + i w^2$, and $z = z^1 + i z^2$, and neglect the remaining four dimensions. We again consider a stack of $N$ D7-branes placed at $f^{(\cancel{A})}(v,w,z)=z=0$. We denote the four real worldvolume coordinates we care about by $\tau, t, \sigma, s$, and choose to work in static gauge: $\tau=v^1$, $t=v^2$, $\sigma=w^1$, $s=w^2$. Given the space-time complex structure chosen, this specifies the complex structure on the worldvolume such that $\tau+it$ and $\sigma+is$ are the holomorphic coordinates. The BPS equations governing this supersymmetric system are written in \eqref{eq:8hitchin}. For the present analysis, however, we will also have to consider the leading $\alpha'$ correction to the non-holomorphic equation, which has been derived in \cite{Minasian:2001na,Butti:2007aq,Marchesano:2010bs,Marchesano:2016cqg}. The latter will indeed be essential for establishing the duality. Written in real coordinates, the full set of equations, including the correction term is:
\begin{equation}\label{eq:hitchin4real}
  \begin{split}
    &D_{\sigma} \phi^8 = D_{s} \phi^9\,,\qquad\ \  D_{\tau} \phi^8 = D_{t} \phi^9\,,\\
    &D_{s} \phi^8 + D_{\sigma} \phi^9 = 0\,,\quad D_{t} \phi^8 + D_{\tau} \phi^9 = 0\,,\\
    &F_{\sigma t} = - F_{s \tau}\,,\qquad\ \ \ \,\,\, F_{\sigma \tau} = F_{st}\,,\\
    & \mathcal{D}:= (F_{\sigma s} + F_{\tau t} + \phi^c) + \tfrac12\lambda^2\epsilon^{\alpha\beta\gamma\delta} ( D_\alpha\phi^8 D_\beta\phi^9 F_{\gamma\delta} - \tfrac{1}{4} \phi^c F_{\alpha\beta}  F_{\gamma\delta} )= 0\,,
  \end{split}
\end{equation}
where, as before, we have defined $\phi^c := -i[\phi^8, \phi^9]$. Recall that we focus on configurations where $\Tr\phi^8=\Tr\phi^9=0$, and where $\phi^c$ is along the Cartan of the gauge group. The last term of the last equation is an $\alpha'^2$ correction to the BPS equations \cite{Marchesano:2010bs,Marchesano:2016cqg}, and to write it we collectively indicated the worldvolume coordinates as $\xi^\alpha =\{\tau, t, \sigma, s\}$, with $\epsilon^{\tau t\sigma s}=1$. For dimensional reasons, and because the D7-brane stack is flat, this is the only non-vanishing $\alpha'$ correction to the BPS equations within the regime of validity of the Hitchin-system description.

We are now in the position to write the general WZ action Eq.~(\ref{eq:WZagen}) for a four-supercharge configuration, which we split into three terms corresponding to the coupling to different RR-potentials:
\begin{equation}
  \begin{split}
    S_{\WZ}^{(\cancel{A})} &= S_{\WZ}^{(\cancel{A})}|_{C_8} + S_{\WZ}^{(\cancel{A})}|_{C_6} + S_{\WZ}^{(\cancel{A})}|_{C_4} = \mu_7 \int \STr\{P[C_8] + i \lambda^2 P[\iota_{\pmb{\phi}} \iota_{\pmb{\phi}} C_8] \w F_2 \} \\
    &\qquad\qquad\qquad + \mu_7 \int \STr\{\lambda P[C_6] \w F_2 + \frac{i}{2} \lambda^3 P[\iota_{\pmb{\phi}} \iota_{\pmb{\phi}} C_6 \w F \w F]\} \\
    &\qquad\qquad\qquad + \mu_7 \frac{\lambda^2}{2} \int \STr\{ P[C_4] \w F_2 \w F_2\}\,.
  \end{split}
\end{equation}
The curvature corrections in \eqref{eq:WZagen} are vanishing here, because the brane worldvolume is flat and trivially embedded. The three terms above correspond to D$7$-, D$5$-, and D$3$-charges respectively, which we are going to consider in order.


\subsubsection*{D7-brane charges}

The terms multiplying $C_8$ can be written as (see \cite{MAT} for details)\footnote{Analogously to the 8-supercharge discussion,  we use the notation: $C_8 = \frac{1}{4!} \sum_{i,j,k,l} C_8^{(X^iX^jX^kX^l)} \w \d X^i\wedge \d X^j\wedge \d X^k\wedge \d X^l$, where $X^i=\{v^1,v^2,w^1,w^2,z^1,z^2\}$ are collective coordinates for the $\mathbb{C}^3$ subspace of the space-time. Note that, in our notation, $C_8^{(X^iX^jX^kX^l)}$ are \emph{four forms}.}

\begin{equation}\label{eq:4naC8}
  \begin{split}
    S_{\WZ}^{(\cancel{A})}|_{C_8} &= \mu_7 \int  \d \tau \w \d t \w \d \sigma \w \d s \w \Big( N C_8^{(v^1 v^2 w^1 w^2 )} \\
    & \qquad\quad + \lambda^2 C_8^{(w^1 w^2 z^1 z^2)} \STr\{(D_\tau \phi^8)^2 + (D_t \phi^8)^2 - \phi^c F_{\tau t}\}\\
    & \qquad\quad + \lambda^2 C_8^{(v^1 v^2 z^1 z^2)} \STr\{(D_\sigma \phi^8)^2 + (D_s \phi^8)^2 - \phi^c F_{\sigma s}\}\\
    & \qquad\quad + \lambda^2 \left(C_8^{(w^1 v^1 z^1 z^2)} + C_8^{(w^2 v^2 z^1 z^2)}\right)\STr\{D_\sigma \phi^8 D_t \phi^8 - D_s \phi^8 D_\tau \phi^8 + \phi^c F_{st}\}\\
    & \qquad\quad + \lambda^2 \left(C_8^{(w^2 v^1 z^1 z^2)} - C_8^{(w^1 v^2 z^1 z^2)}\right)\STr\{D_\sigma \phi^8 D_\tau \phi^8 + D_s \phi^8 D_t \phi^8 + \phi^c F_{s\tau}\}\Big)\,,
  \end{split}
\end{equation}
where linear terms in $\phi$ have been removed because of their tracelessness, and all but the last equation in \eqref{eq:hitchin4real} have been used to get rid of $\phi^9$ and to simplify the expression.


\subsubsection*{D5-brane charges}

The terms multiplying $C_6$ can be written as (see \cite{MAT} for details)
\begin{equation}\label{eq:naC6}
  \begin{split}
    S_{\WZ}^{(\cancel{A})}|_{C_6} &= \mu_7 \lambda \int \d \tau \w \d t \w\d \sigma \w \d s \w  \STr\Big\{ \\
      & C_6^{(w^1 w^2)} F_{\tau t} + C_6^{(v^1 v^2)} F_{\sigma s} - (C_6^{(w^1 v^1)} + C_6^{(w^2 v^2)})F_{st} + (C_6^{(w^1 v^2)} - C_6^{(w^2 v^1)}) F_{s\tau}\\
      & \qquad + \lambda (C_6^{(v^1 z^1)} + C_6^{(v^2 z^2)}) (D_t \phi^8 F_{\sigma s} + D_\sigma \phi^8 F_{st} + D_s \phi^8 F_{s\tau})\\
      & \qquad - \lambda (C_6^{(v^2 z^1)} - C_6^{(v^1 z^2)}) (D_\tau \phi^8 F_{\sigma s} - D_s \phi^8 F_{st} + D_\sigma \phi^8 F_{s\tau})\\
      & \qquad - \lambda (C_6^{(w^1 z^1)} + C_6^{(w^2 z^2)}) (D_s \phi^8 F_{\tau t} - D_\tau \phi^8 F_{st} + D_t \phi^8 F_{s\tau})\\
      & \qquad - \lambda (C_6^{(w^2 z^1)} - C_6^{(w^1 z^2)}) (D_\sigma \phi^8 F_{\tau t} + D_t \phi^8 F_{st} + D_\tau \phi^8 F_{s\tau})\\
      & \qquad + \lambda^2 C_{6}^{(z^1 z^2)}\epsilon^{\alpha\beta\gamma\delta} ( D_\alpha\phi^8 D_\beta\phi^9 F_{\gamma\delta} - \tfrac{1}{4} \phi^c F_{\alpha\beta}  F_{\gamma\delta} ) \Big\}\,,
  \end{split}
\end{equation}
where, as usual, we have been using all but the last equation in \eqref{eq:hitchin4real} to simplify the expression. The coefficient of $C_{6}^{(z^1 z^2)}$, is proportional to the $\lambda^2$ correction to the Hitchin system, written in the last line of \eqref{eq:hitchin4real}. None of these terms vanishes in general. In particular, the first line is generically non-zero because, as opposed to the eight-supercharge case, $\Tr\phi^c=0$ does not imply the tracelessness of $F_2$. We will come back to the structure of the induced D5-brane charges in Section \ref{ssec:dual4}, where we will focus on configurations with a vanishing trace of all components of $F_2$.


\subsubsection*{D3-brane charges}

The terms multiplying $C_4$ can be written as (see \cite{MAT} for details)
\begin{equation}\label{eq:nAD3WZ}
  \begin{split}
    S_{\WZ}^{(\cancel{A})}|_{C_4} &= \mu_7\lambda^2 \int \d \tau \w \d t \w \d \sigma \w \d s \w C_4\, \Tr\{F_{\tau t} F_{\sigma s} - (F_{st}^2 + F_{s \tau}^2)\}   \,,
  \end{split}
\end{equation}
where we have used the equations in the third line of \eqref{eq:hitchin4real}.
Note that the above expression does not have a definite sign, because in T-brane configurations $F_2$ is not anti-self dual. In contrast, for ordinary D-brane solutions, $\phi^c$ vanishes. For low gradients of $\Phi$ in string units, we can neglect the $\lambda^2$ correction in \eqref{eq:hitchin4real}, which implies primitivity of $F_2$: $F_2 \w J=0$. This in turn leads to $F_{\tau t}=-F_{\sigma s}$, making the above expression a sum of three perfect squares and thus of negative definite sign.


\subsubsection*{DBI  action}

The computation of the non-Abelian DBI  action is significantly more involved. The quantity we would like to compute is \eqref{eq:DBI} by imposing that the system satisfies the set of equations \eqref{eq:hitchin4real}. Since we are in flat space, \eqref{eq:DBI} simplifies to \cite{Myers:1999ps}:
\begin{align}
S ^{(\cancel A)}_{\DBI}=-\mu_7  \int \d^8 \xi e^{-\phi} \text{STr} \left[\sqrt{-\text{det}\left(\eta_{\alpha \beta} + \lambda^2 D_\alpha \phi^a Q_{ab}^{-1} D_\beta \phi^b + \lambda F_{\alpha \beta}\right) \det Q}\right]\,.
\end{align}
Remarkably, it is possible to show that, after imposing all but the last equation in \eqref{eq:hitchin4real}, the DBI  action significantly simplifies leaving the expression
\begin{align}\label{eq:DBI4na}
S ^{(\cancel A)}_{\DBI}=-\mu_7  \int \d^8 \xi e^{-\phi} \text{STr}\left[\sqrt{\mathcal T^2 + \lambda^2 \mathcal D^2}\right]\,,
\end{align}
where we defined the quantity $\mathcal T$ as
\begin{align}
\mathcal T :=&\;1 + \lambda^2  \left[ (D_{\sigma} \phi^8)^2 + (D_{s} \phi^8)^2 + (D_{\tau} \phi^8)^2 + (D_{t} \phi^8)^2 \right.\\
&\left.      - \phi^c (F_{\sigma s} + F_{\tau t}) + (F_{st}^2 + F_{s \tau}^2 - F_{\sigma s} F_{\tau t}) \right]\,,
\end{align}
whereas $\mathcal D$ is defined in the last line of \eqref{eq:hitchin4real}. Supersymmetry requires $\mathcal D=0$, implying that the argument of the square root in \eqref{eq:DBI4na} be a perfect square, as expected. This highly non-trivial simplification confirms the results of \cite{Marchesano:2010bs,Marchesano:2016cqg} on the form of the $\alpha'$ corrections to the BPS equations. As mentioned in the introduction, the form of the DBI we are using is known to deviate from string perturbation theory starting at order $\alpha'^3$, while we are limiting ourselves to a study including corrections up to  $\alpha'^2$ order. Since no deviation is present at this order, and the DBI takes the form \eqref{eq:DBI4na} on shell, this makes us confident that our truncation to order $\alpha'^2$ is consistent.

Summarizing we find that the DBI action becomes (see \cite{MAT} for details)
\begin{equation}\label{eq:4nADBI}
  \begin{split}
    S^{(\cancel{A})}_{\DBI} & = - \mu_7 \int e^{-\phi} \d^4\xi \d \tau \d t \d \sigma \d s \Big| N + \lambda^2 \Tr \Big\{ (D_{\sigma} \phi^8)^2 + (D_{s} \phi^8)^2 + (D_{\tau} \phi^8)^2 + (D_{t} \phi^8)^2 \\
      &\qquad\qquad\qquad - \phi^c (F_{\sigma s} + F_{\tau t}) + (F_{st}^2 + F_{s \tau}^2 - F_{\sigma s} F_{\tau t}) \Big\} \Big|\,.
  \end{split}
\end{equation}


\subsection{The Abelian side}\label{ssec:4A}

On the Abelian side we can no longer a priori expect the T-brane physics to be captured by curvature only. The T-brane will generically have lower-dimensional brane charges induced by the terms derived in the previous section (cf.~Equations \eqref{eq:naC6} and \eqref{eq:nAD3WZ}), and the Abelian brane may need a worldvolume flux switched on in order to induce these charges.

We still look for the same structure as for the branes preserving eight supercharges, expecting that the $N$ one-dimensional K\"ahler strands of Abelian branes now are $N$ two-dimensional K\"ahler sheets. The goal is to write down the duality conditions to be satisfied by the holomorphic embedding $f^{(A)}(v,w,z) = 0$ of the Abelian brane. As before, we find it more convenient to invert $f^{(A)}$, and formulate the expressions in terms of $z = z(v,w)$.

Let us start by focusing on the WZ action \eqref{eq:WZagen}. The terms we need to consider are
\begin{equation}
  \begin{split}
    S_{\WZ}^{({A})} &= S_{\WZ}^{({A})}|_{C_8} + S_{\WZ}^{({A})}|_{C_6} + S_{\WZ}^{({A})}|_{C_4} \\
            &= \mu_7 \int P[C_8] + \lambda P[C_6] \w F_2 + C_4 \w \left(\frac{\lambda^2}{2} F_2 \w F_2 + \left. \sqrt{\frac{\hat{\mathcal{A}}(2\pi \lambda R_T)}{\hat{\mathcal{A}}(2\pi \lambda R_N)}} \right|_{4} \right)\,.
  \end{split}
\end{equation}
This is the action for a single sheet, and the matching with the non-Abelian WZ action is performed as before by considering $N$ sheets with different phases adding up to zero.
We will now discuss the various induced brane charges.


\subsubsection*{D7-brane charges}

The terms multiplying $C_8$ can be written as (see \cite{MAT} for details)
\begin{equation}\label{eq:4aC8}
  \begin{split}
    S_{\WZ}^{({A})}|_{C_8} &= \mu_7 N \int \d \tau \w \d t \w \d \sigma \w \d s \w  \det \mathcal{J} \Big(C_8^{( v^1 v^2 w^1 w^2)} + C_8^{(w^1 w^2 z^1 z^2)} |\partial_v z|^2 + C_8^{(v^1 v^2 z^1 z^2)} |\partial_w z|^2 \\
    &\qquad + (C_8^{(w^2 v^1 z^1 z^2)} - C_8^{(w^1 v^2 z^1 z^2)}) \text{Re}(\partial_w z \partial_{\bar{v}} \bar{z}) + (C_8^{(w^1 v^1 z^1 z^2)} + C_8^{(w^2 v^2 z^1 z^2)}) \text{Im}(\partial_w z \partial_{\bar{v}} \bar{z}) \Big) \,,
  \end{split}
\end{equation}
where $\mathcal{J} = [\partial X^i/\partial \xi^\alpha]$ is the Jacobian matrix, with $X^i =\{ v^1, v^2, w^1, w^2\}$, $\xi^\alpha =\{\tau, t, \sigma, s\}$. Here, for simplicity, in writing the sum over the $N$ sheets, we ignored all terms linear in $z$ ($\bar z$), which are canceling under the sum.


\subsubsection*{D5-brane charges}

The terms multiplying $C_{6}$ can be expressed using the following six quantities
\begin{equation}\label{eq:alphadef}
  \alpha^{ij} := - \frac{1}{2} \epsilon^{\alpha \beta \gamma \delta} \mathcal{J}^i_{\alpha} \mathcal{J}^j_{\beta} F_{\gamma \delta}\,,
\end{equation}
and we obtain (see \cite{MAT} for details):
\begin{equation}\label{eq:aC6}
  \begin{split}
    S&{}^{(A)}_{\WZ}|_{C_{6}} = \lambda \mu_7 \int \d \tau \w \d t \w \d \sigma \w \d s \w \Big[ \\
    & C_6^{(v^1 v^2)} \alpha^{v^1 v^2} + C_6^{(v^1 w^1)} \alpha^{v^1 w^1} + C_6^{(v^1 w^2)} \alpha^{v^1 w^2} + C_6^{(v^2 w^2)} \alpha^{v^2 w^2} + C_6^{(w^1 w^2)} \alpha^{w^1 w^2} + C_6^{(v^2 w^1)} \alpha^{v^2 w^1}\\
    & - C_6^{(v^1 z^1)} (\text{Im}(\partial_v z) \alpha^{v^1 v^2} - \text{Re}(\partial_w z) \alpha^{v^1 w^1} + \text{Im}(\partial_w z) \alpha^{v^1 w^2}) \\
    & + C_6^{(v^1 z^2)} (\text{Re}(\partial_v z) \alpha^{v^1 v^2} + \text{Im}(\partial_w z) \alpha^{v^1 w^1} + \text{Re}(\partial_w z) \alpha^{v^1 w^2})  \\
    & - C_6^{(v^2 z^1)} (\text{Re}(\partial_v z) \alpha^{v^1 v^2} - \text{Re}(\partial_w z) \alpha^{v^2 w^1} + \text{Im}(\partial_w z) \alpha^{v^2 w^2}) \\
    & - C_6^{(v^2 z^2)} (\text{Im}(\partial_v z) \alpha^{v^1 v^2} - \text{Im}(\partial_w z) \alpha^{v^2 w^1} - \text{Re}(\partial_w z) \alpha^{v^2 w^2}) \\
    & - C_6^{(w^1 z^1)} (\text{Re}(\partial_v z) \alpha^{v^1 w^1} - \text{Im}(\partial_v z) \alpha^{v^2 w^1} + \text{Im}(\partial_w z) \alpha^{w^1 w^2})\\
    & - C_6^{(w^2 z^2)} (\text{Im}(\partial_v z) \alpha^{v^1 w^2} + \text{Re}(\partial_v z) \alpha^{v^2 w^2} + \text{Im}(\partial_w z) \alpha^{w^1 w^2}) \\
    & - C_6^{(w^2 z^1)} (\text{Re}(\partial_v z) \alpha^{v^1 w^2} - \text{Im}(\partial_v z) \alpha^{v^2 w^2} + \text{Re}(\partial_w z) \alpha^{w^1 w^2}) \\
    & - C_6^{(w^1 z^2)} (\text{Im}(\partial_v z) \alpha^{v^1 w^1} + \text{Re}(\partial_v z) \alpha^{v^2 w^1} - \text{Re}(\partial_w z) \alpha^{w^1 w^2}) \\
    & + C_6^{(z^1 z^2)} \Big(\text{Im}\Big[\partial_w z \partial_{\bar{v}} \bar{z}  ((\alpha^{v^1 w^1} + \alpha^{v^2 w^2}) + i (\alpha^{v^1 w^2} - \alpha^{v^2 w^1}))\Big] + |\partial_v z|^2 \alpha^{v^1 v^2}  + |\partial_w z|^2 \alpha^{w^1 w^2}\Big)\Big]\,.
  \end{split}
\end{equation}
This expression is for a single Abelian sheet, and no phase-shifted sum has been carried out yet. The reason is that we can in principle allow a different $F_2$ on each sheet. We postpone this discussion to Section \ref{ssec:dual4}.


\subsubsection*{D3-brane charges}

To write the terms multiplying $C_{4}$ we need to evaluate the $\hat{\mathcal{A}}$-roof genus in \eqref{eq:WZagen}. The latter can be expanded in terms of Pontryagin classes for both the tangent and the normal bundle of the brane worldvolume $S$
\begin{equation}
  \sqrt{\frac{\hat{\mathcal{A}}(2\pi \lambda R_T)}{\hat{\mathcal{A}}(2\pi \lambda R_N)}} = 1 - \frac{(2\pi \lambda)^2}{48}(p_1(TS) - p_1(NS)) + \ldots \,,
\end{equation}
where the ellipsis stand for higher-form terms, which vanish for dimensional reasons.
The Pontryagin classes can in turn be written in terms of Chern classes. Indeed we have that for any vector bundle $E$ carrying a complex structure the first Pontryagin class may be written as
\begin{align}
p_1(E) = c_1^2(E) -2 c_2(E)\,.
\end{align}
Moreover in our case the normal bundle $NS$ is a line bundle and therefore its second Chern class vanishes. Using the fact that the brane wrapping $S$ is embedded in flat space, we can use adjunction and get
\begin{align}\label{eq:adjunction}
c(TS) \wedge c(NS)=0\,,
\end{align}
implying, in particular, that $c_1(TS)+c_1(NS) = 0$.\footnote{Using the second fundamental form (which we will introduce shortly), it is possible to show that at the level of representatives of cohomology classes this relationship is $\text{tr} [R_T] = -\text{tr} [R_N]$.} This allows us to write the induced D3-brane charge as
\begin{align}\label{eq:AD3WZ}
S^{(A)}_\WZ|_{C_4} &= \mu_7 \int C_4 \wedge (2\pi \lambda)^2 \left(\frac{1}{8 \pi^2}F_2 \wedge F_2 -\frac{1}{24}c_2(TS)\right)\,.
\end{align}
For later comparison we will find it useful to write the second Chern class in terms of the curvature two form:
\begin{align}
c_2(TS) = \frac{1}{8 \pi^2} \left(\Tr[R_T \wedge R_T] - [\Tr R_T]\wedge [\Tr R_T]\right)\,.
\end{align}
%


\subsubsection*{DBI  action}

On the Abelian side of this four-supercharge system the $\alpha'$-corrected DBI action \eqref{eq:DBI} takes the following form
\begin{align}\label{eq:DBI4s}
S^{(A)}_\DBI = -\mu_7 \int d^8 \xi \, e^{-\phi} \sqrt{-g}&\left[1- \frac{1}{24} \frac{(2\pi \lambda)^2}{32\pi^2} \left( (R_T)_{\alpha\beta\gamma\delta}(R_T)^{\alpha\beta\gamma\delta} - 2(R_T)_{\alpha\beta}(R_T)^{\alpha\beta}\right.\right.\nonumber \\& \left.\left.- (R_N)_{\alpha\beta ab}(R_N)^{\alpha \beta ab} + 2 \bar{R}_{ab} \bar{R}^{ab}\right) \right]\,,
\end{align}
where $g$ is the determinant of the pull-back of the flat metric onto the brane worldvolume, and, for simplicity, we have chosen to work with vanishing worldvolume flux\footnote{We will see in Section \ref{ssec:dual4} that this choice will not impede the duality matching.}.

We now need to compute the curvature corrections as well as the determinant of the metric. In order to do so, we need to obtain the expressions for the various combinations of the Riemann tensor appearing in \eqref{eq:DBI4s}. We can use the Gau\ss--Codazzi equations to rewrite them in terms of the Riemann tensor $R$ of the ambient space and the second fundamental form $\Omega$ (see for instance \cite{Bachas:1999um} for further details):
\begin{align}
(R_T)_{\alpha  \beta \gamma  \delta} &= R_{\alpha  \beta \gamma  \delta} + \delta_{a  b} \left(\Omega^a_{\alpha \gamma} { \Omega}^{ b}_{ \beta  \delta}-\Omega^a_{\alpha \delta} { \Omega}^{ b}_{ \beta  \gamma}\right)\,,\\
(R_N)_{\alpha  \beta}{}^{a  b} &=  R^{a  b}{}_{\alpha  \beta} +g^{\gamma \delta}\left( \Omega^a_{\alpha \gamma} { \Omega}^{ b}_{\beta \delta}- \Omega^b_{\alpha \gamma} { \Omega}^{ a}_{\beta \delta}\right)\,,\\
\bar R_{a  b} &= R^\alpha{}_{ a \alpha  b}+ g^{\alpha \gamma} g^{\beta \delta} \Omega_{a| \alpha \beta} \Omega_{b|\gamma \delta}\,.
\end{align}
Recall that greek (latin) indices denote directions tangent (normal) to the embedded manifold. Moreover we called $g_{\alpha \beta}$ the metric on the tangent bundle and we have the trivial metric $\delta_{ab} $ on the normal bundle. The second fundamental form may be written in terms of the embedding functions $X^a=X^a(\xi^\alpha)$ as
\begin{align}
  \Omega^a_{\alpha \beta} = \partial_\alpha \partial_\beta X^a - (\Gamma_T)^\gamma_{\ \alpha \beta} \partial_\gamma X^a + \Gamma^a_{\ bc} \partial_\alpha X^b \partial_\beta X^c\,,
\end{align}
where $\Gamma_T$ is the Levi-Civita connection built using the metric on the tangent bundle of the embedded manifold, and $ \Gamma$ is the Levi-Civita connection of the ambient space. In our case two kinds of simplification occur: First the ambient space is flat, so that $R=0, \,\Gamma=0$, and the brane worldvolume is a K\"ahler manifold embedded holomorphically. This greatly simplifies the structure of the second fundamental form: Passing to complex indices, but keeping the same symbols for them, the only two possible tensor structures are $\Omega^a_{\alpha \beta}$ and its conjugate $\overline \Omega^{\bar a}_{\bar \alpha \bar \beta}$. It is now simple to show that the following relations hold
\begin{align}
&(R_T)_{\alpha \bar \beta \gamma \bar \delta} (R_T)^{\alpha \bar \beta \gamma \bar \delta} = - \bar R_{a \bar b} \bar R^{a \bar b}\,,\\
&(R_N)_{\alpha \bar \beta a \bar b} (R_N)^{\alpha \bar \beta a \bar b} = - (R_T)_{\alpha \bar \beta} (R_T)^{\alpha \bar \beta}\,.
\end{align}
Therefore \eqref{eq:DBI4s} simplifies to
\begin{align}
S^{(A)}_\DBI &=- \mu_7 \int d^8 \xi \, e^{-\phi} \sqrt{-g}\left[1+\frac{1}{24} \frac{(2\pi \lambda)^2}{16\pi^2} \left( (R_T)_{\alpha\bar \beta\gamma\bar\delta}(R_T)^{\alpha\bar \beta\gamma\bar \delta} + (R_T)_{\alpha\bar \beta}(R_T)^{\alpha\bar \beta}\right)\right]\nonumber\\
&=-\mu_7 \int d^4 \xi \, e^{-\phi} \left[\sqrt{-g} \d \tau \d t \d \sigma \d s -\frac{1}{24} \frac{(2\pi \lambda)^2}{8\pi^2} \Big( \Tr[R_T\wedge \star R_T] - \Tr[R_T]\wedge \star \Tr[R_T]\Big)\right]\,.
\end{align}
The part that remains to be computed is the determinant of the metric. Using the condition that the embedding is holomorphic, the argument of the square root becomes a perfect square giving the simple expression (see \cite{MAT} for details)
\begin{align}
\sqrt{-g} = \left(1 + |\partial_v z|^2 + |\partial_w z|^2\right)\det \mathcal J\,.
\end{align}
Combining both results, we find that the DBI  action is
\begin{align}\label{eq:4ADBI}
    S^{(A)}_{\DBI}  = - \mu_7 N \int& \d^4\xi e^{-\phi} \Big[ \d \tau \d t\d \sigma \d s \left(1 + |\partial_v z|^2 + |\partial_w z|^2 \right) \det \mathcal{J} \nonumber\\
    &\qquad\qquad -\frac{1}{12} \frac{(2\pi \lambda)^2}{16\pi^2} \left( \Tr[R_T\wedge \star R_T] - \Tr[R_T]\wedge \star \Tr[R_T]\right)\Big]\,,
\end{align}
where we have also taken the sum over all sheets to produce the overall factor of $N$.


\subsection{Duality dictionary}\label{ssec:dual4}

Analogously to the eight-supercharge T-branes discussed in Section \ref{sssec:dd8}, we now proceed to compare the two pairs of actions derived in the Sections \ref{ssec:4nA} and \ref{ssec:4A} for generic space-time data, and establish in this way our brane-brane duality.


\subsubsection*{D7-brane charges}

Equating the non-Abelian with the Abelian WZ action at the level of $C_8$ terms, \eqref{eq:4naC8} and \eqref{eq:4aC8} respectively, leads to the following conditions:
\begin{equation}\label{eq:4D7match}
  \begin{split}
    \det(\mathcal{J}) &= 1\,, \\
    |\partial_w z|^2 &= \frac{\lambda^2}{N} \Tr\{(D_\sigma \phi^8)^2 + (D_s \phi^8)^2 - \phi^c F_{\sigma s}\}\,,\\
    |\partial_v z|^2 &= \frac{\lambda^2}{N} \Tr\{(D_\tau \phi^8)^2 + (D_t \phi^8)^2 - \phi^c F_{\tau t}\}\,,\\
    \partial_w z \partial_{\bar{v}} \bar{z} &= \frac{\lambda^2}{N} \Tr \{ D_\sigma \phi^8 D_t \phi^8 - D_s \phi^8 D_\tau \phi^8 + \phi^c F_{st} \}\,,\\
    &\quad + \frac{\lambda^2}{N} i \Tr \{ D_\sigma \phi^8 D_\tau \phi^8 - D_s \phi^8 D_t \phi^8 + \phi^c F_{s\tau} \}\,.
  \end{split}
\end{equation}
Note the appearance of two extra equations compared to the eight-supercharge system: The third equation, which is just the analog of the second one for the extra complex coordinate $v$, and the fourth equation, which mixes the two complex directions $v$ and $w$. We can recover the eight-supercharge system of duality conditions \eqref{eq:8duality1} and \eqref{eq:8duality2} by simply putting $\partial_v z = 0$, and removing all objects with $\tau, t$ indices. For standard intersecting D7-brane configurations (namely with $\phi^c=0$) the above set of equations is solved by an identification analogous to \eqref{eq:8duality2nonTbis}, where we now allow dependence also on $v$.


\subsubsection*{D5-brane charges}

Equating now the two WZ actions at the level of $C_6$ terms, \eqref{eq:naC6} and \eqref{eq:aC6} respectively, gives us extra duality conditions. First let us consider the components of $C_{6}$ without legs along $z$, which appear in the first lines of both \eqref{eq:naC6} and \eqref{eq:aC6}. On the non-Abelian side these terms contain a trace of a single flux component, and they are supposed to be related to the $\alpha$ coefficients (which contain the Abelian flux $F_2^{(A)}$) as follows:\footnote{We do not use the equal sign here, because in Equation \eqref{eq:aC6} we have not performed the sum over all the $N$ Abelian sheets.}
\begin{equation}\label{eq:AvsNAC6}
  \begin{split}
    \alpha^{w^1 w^2} \longleftrightarrow \Tr F_{\tau t}^{(\cancel{A})}\,,\qquad    &  \alpha^{w^1 v^1},\alpha^{w^2 v^2} \longleftrightarrow - \Tr F_{st}^{(\cancel{A})}\,,\\
    \alpha^{v^1 v^2} \longleftrightarrow \Tr F_{\sigma s}^{(\cancel{A})}\,,\qquad  &  \alpha^{w^1 v^2},-\alpha^{w^2 v^1} \longleftrightarrow \Tr F_{s\tau}^{(\cancel{A})}\,.\\
  \end{split}
\end{equation}
The D$5$-brane charges induced by these non-Abelian-flux traces are not really genuine T-brane data, because they contain no information about the commutator of $\phi$'s. In other words, they only ``dress'' the T-brane data with extra information which does not affect the non-Abelian behavior of the system. Therefore, to keep the present discussion as simple as possible, we focus on ``pure'' T-brane configurations, namely those having \emph{every} component of $F_2^{(\cancel{A})}$ traceless. In particular, since we want the non-Abelian features of our configuration to only originate from $\Phi$, we take \emph{every} component of the worldvolume flux to lie along the Cartan of the gauge group.

To determine the Abelian dual of these types of T-brane configurations, Equation (\ref{eq:AvsNAC6}) suggests us that we can take $\alpha^{ij}=0\;\;\forall i,j$. From the definition \eqref{eq:alphadef}, the easiest possible solution for the Abelian flux is then $F_{\alpha\beta}^{(A)}=0\;\;\forall\alpha,\beta$. Given the homogeneity of Eq.~(\ref{eq:aC6}) in the $\alpha$ parameters, setting all of them to zero removes all induced D5-brane charges. This is incompatible with the duality, unless all terms in Eq.~(\ref{eq:naC6}) are vanishing too, for the class of T-branes we are considering. Remarkably, this is exactly what happens. Indeed, the terms linear in $\lambda$ vanish because we have set $\Tr F_{\alpha\beta}^{(\cancel{A})}=0\;\;\;\forall\alpha,\beta$. The terms quadratic in $\lambda$ vanish for a simple group-theory reason: Since all $F_{\alpha\beta}^{(\cancel{A})}$ are taken to lie in the Cartan subalgebra, whereas all $D_\alpha\phi^8$ are linear combinations of root generators\footnote{This is true because $[\Phi,\Phi^\dagger]$ is along the Cartan, and this implies that $\Phi$ can only be a linear combinations of the generators associated to all simple roots and to the negative of the highest root (modulo Weyl reflections).}, their products are always traceless, because the Killing form does not mix Cartan with root directions. Finally, the term cubic in $\lambda$ is zero because of supersymmetry: Using the tracelessness of all flux components, it coincides with $\Tr\mathcal{D}$, with $\mathcal{D}$ defined in the last line of \eqref{eq:hitchin4real}.

To summarize, the sort of ``pure'' T-brane configurations we focus on here admit an Abelian dual made of curved D7-branes with vanishing worldvolume flux, exactly like in the eight-supercharge systems discussed in Section \ref{sec:8susies}. Note that extending the duality to non-Abelian configurations in which the T-brane is not pure, namely has non-trivial D5-brane charge induced, does necessitate $F_2^{(A)} \neq 0$ on the Abelian side, but we will not consider such configurations here.


\subsubsection*{D3-brane charges}

The last duality condition comes from requiring the matching of the $C_4$ couplings (the induced D$3$-brane charges), namely Eq. \eqref{eq:nAD3WZ} with Eq. \eqref{eq:AD3WZ}. Using that the Abelian side has vanishing worldvolume flux, this leads to
\begin{equation}\label{eq:4D3match}
  \frac{N}{48} \Big(\Tr[R_T \wedge R_T] - [\Tr R_T]\wedge [\Tr R_T]\Big) =\Tr \left\{ F^2_{st} + F^2_{s\tau} - F_{\sigma s} F_{\tau t} \right\}\d \tau \w \d t \w \d \sigma \w \d s\,,
\end{equation}
where the curvature of the fluxless Abelian brane appears on the l.h.s., and the flux of the flat non-Abelian stack appears on the r.h.s.. To summarize, once an embedding is found that solves the system (\ref{eq:4D7match}), the curvature induced by that embedding is also subject to the above condition.


\subsubsection*{DBI  action}

Finally, we need to check the matching of the two DBI  actions: Equation (\ref{eq:4nADBI}) on the non-Abelian side and Equation (\ref{eq:4ADBI}) on the Abelian side.
For configurations preserving eight supercharges, analyzed in Section \ref{sec:8susies}, we showed that the matching of the DBI  actions follows from the matching of the WZ actions, without imposing further constraints. Here, on the contrary, we will get an extra consistency condition.

For the reader's convenience, we rewrite the two DBI  actions (\ref{eq:4nADBI}) and (\ref{eq:4ADBI}) here
\begin{equation}
  \begin{split}
    S^{(\cancel{A})}_{\DBI} = - \mu_7 \int& e^{-\phi} \d^4\xi \d \tau \d t \d \sigma \d s \Big| N + \lambda^2 \Tr \Big\{ (D_{\sigma} \phi^8)^2 + (D_{s} \phi^8)^2 + (D_{\tau} \phi^8)^2 + (D_{t} \phi^8)^2 \\
      &\qquad\qquad\qquad - \phi^c (F_{\sigma s} + F_{\tau t}) + (F_{\sigma \tau}^2 + F_{\sigma t}^2 - F_{\sigma s} F_{\tau t}) \Big\} \Big|\,,\\
      S^{(A)}_{\DBI}  =  -\mu_7 N \int& \d^4\xi e^{-\phi} \Big[ \d \tau \d t \d \sigma \d s \left(1 + |\partial_v z|^2 + |\partial_w z|^2 \right) \det \mathcal{J} \nonumber\\
      &\qquad\qquad -\frac{1}{12} \frac{(2\pi \lambda)^2}{16\pi^2} \left( \Tr[R_T\wedge \star R_T] - \Tr[R_T]\wedge \star \Tr[R_T]\right)\Big]\,.
  \end{split}
\end{equation}
Using Equations \eqref{eq:4D7match} and \eqref{eq:4D3match}, we can readily see that the above actions do not match, unless the curvature of the Abelian brane satisfies the following condition
\begin{equation}\label{eq:CurvatureConstraint}
  \Tr[R_T\wedge \star R_T] - \Tr[R_T]\wedge \star \Tr[R_T] =- \Tr[R_T\wedge  R_T] + \Tr[R_T]\wedge \Tr[R_T]\,.
\end{equation}
This constraint could be easily satisfied by imposing anti-selfduality of the curvature two-form, $\star R_T=-R_T$. However, this is too strong, because it would completely kill the curvature. Indeed, writing the adjunction formula \eqref{eq:adjunction} at the level of the four-form representatives, using the second fundamental form, one obtains $\Tr[R_T\w R_T] = -\Tr[R_T] \w \Tr [R_T]$. Anti-selfduality for $R_T$ would imply that the l.h.s.~of this equation is non-positive while the r.h.s.~is non-negative, thereby constraining $R_T$ to identically vanish.

It is not clear to us whether the curvature constraint \eqref{eq:CurvatureConstraint} follows from the form of the embedding that solves \eqref{eq:4D7match}, or it imposes further conditions on it. In other words, it would be interesting to understand to what extent the set of duality equations \eqref{eq:4D7match} fixes the form of the Abelian-brane embedding. We will further elaborate on this open problem in the next section.


\section{Discussion}\label{sec:Discussion}

In this work we have found what we believe to be substantial evidence for the proposal made in \cite{Bena:2016oqr}, that the physics of T-branes, in the regime of parameters where the expectation values of the worldvolume fields are larger than the string scale, is described by multiple Abelian branes wrapping curved holomorphic manifolds. This duality makes T-branes much less mysterious from a brane-model-building perspective.

The evidence for this duality presented in \cite{Bena:2016oqr} was obtained in the large-$N$ limit using a very specific example of a T-brane preserving eight supercharges, and came via a complicated chain of dualities. In contrast, the analysis of this paper applies to {\em all} T-brane solutions, and establishes this duality for finite $N$, by working in only one duality frame. We confirm this by matching the Wess-Zumino and the Dirac-Born-Infeld action of the two dual descriptions. This results in several differential equations, whose solution gives the holomorphic embedding of the Abelian brane dual to a given T-brane configuration. These matching equations are \eqref{eq:8duality1} and \eqref{eq:8duality2} for T-brane profiles preserving eight supercharges, and \eqref{eq:4D7match} for T-branes preserving four supercharges. The latter T-branes can also have a non-trivial D3 charge, and matching this charge imposes a quadratic constraint on the curvature two-form of the Abelian brane \eqref{eq:CurvatureConstraint}, which we believe deserves a deeper understanding.

We have then used the general, algorithmic, methodology we developed for constructing Abelian descriptions of T-branes on the particular configuration analyzed in \cite{Bena:2016oqr}. We showed that the Abelian-brane curve obtained in \cite{Bena:2016oqr} is a solution of the matching equations, and argued that the Abelian description of that T-brane solution consists of a bouquet of Abelian D7 branes (each solving the matching equations) meeting at two points. The analysis of the full space of solutions to the matching equations and the precise selection among these solutions of the Abelian brane embedding dual to a given T-brane profile are interesting directions of further study.

Our work opens up several other interesting directions of investigation. One is to produce more explicit examples of this T-brane duality, possibly with four preserved supercharges and involving a non-constant profile of the Higgs field in the holomorphic gauge.\footnote{The holomorphic gauge is one in which $A^{(0,1)}$ is fixed to zero, thereby making the Higgs field simply holomorphic. Such a gauge is linked to the physical unitary gauge used in this paper by a transformation of the complexified version of the gauge group. See \cite{Cecotti:2010bp} for details.} As alluded to already in the introduction, another direction would be to go beyond the kinematical level at which we have established this duality to determine if this is a true duality of string theory.

When the holomorphic-gauge Higgs field $\Phi^{\rm h}$ is constant and, say, has the form of a $N\times N$ Jordan block, it is pretty straightforward to characterize qualitatively the dual Abelian branes. This is essentially because the trace structure of the right-hand-side of \eqref{eq:8duality2} (or of \eqref{eq:4D7match}) is that of the generalized Pauli matrices, and the space dependence of $\Phi$ factorizes (since it is possible to express it in terms of a single real function). As a consequence, the Abelian brane dual has $N$ D7-brane strands (or sheets), all embedded in the same way and only differing by a phase equal to an $N^{\rm th}$ root of unity. This is analogous to our eight-supercharge example \eqref{eq:embed}, though in general the common profile may be different from the simple $z\sim1/w$ profile. It is crucial to note that even though these appear to be $N$ independent D7-branes, their relative positions are all frozen by the boundary conditions, and hence they behave as a single brane whose only massless degree of freedom is the center-of-mass. This is compatible with the gauge symmetry left unbroken by the $\Phi$-profile of the starting T-brane configuration, which is a single $U(1)$, corresponding also to the center-of-mass motion of the original non-Abelian stack of $N$ D7-branes. Generalizing this story to smaller nilpotent orbits of $SU(N)$ (corresponding to different Jordan-block patterns for $\Phi^{\rm h}$) is obvious, leading to free relative displacements of groups of strands (or sheets) and hence to a larger unbroken symmetry.

A more complicated family of T-branes are those for which some of the entries of $\Phi^{\rm h}$ have non-empty vanishing loci, or whose Higgs field in the holomorphic gauge is non-constant. Since we do not know of any explicit solution in this family at present, we do not have a global picture of what the Abelian dual looks like. However, we can still qualitatively predict what is going to happen patch-wise. Indeed, suppose $\Phi^{\rm h}$ is expanded along the simple roots of $SU(N)$, with generic holomorphic polynomials as coefficients. In the local patch where none of these coefficients vanish\footnote{The region where a holomorphic polynomial does not vanish is an open set, and here we are considering the intersection of all these open sets.}, the Abelian-dual description will have the same structure of strands (or sheets) as the T-branes with a constant maximal Jordan block discussed above. However, when we restrict to the local patch of the complex codimension-one locus where one of the holomorphic polynomials vanishes while all the others are non-zero\footnote{For configurations preserving eight supercharges, all the other polynomials will generically be non-zero over this locus.}, the size of the maximal Jordan block splits into two, the unbroken symmetry enhances, and consequently the Abelian dual will be characterized by two independent groups of strands (or sheets). For configurations with four supercharges, there will generically be also a complex codimension-two locus, where two holomorphic polynomials vanish simultaneously and the symmetry enhances further: There, the Abelian dual will be characterized by three independent groups of sheets. Therefore, we can always describe the branch structure of the Abelian dual of a general nilpotent T-brane by patching together different Jordan-block patterns. It would be very interesting to work out, in a given example, the global profile of a multi-branched D7-brane which interpolates between these different patch-wise descriptions. Because of this varying branching structure, we expect such a global profile to involve non-constant phase shifts.

Another class of T-brane configurations whose Abelian dual would be very interesting to construct is the one involving non-trivial monodromies \cite{Cecotti:2010bp}. Again, let us give here a qualitative picture, leaving a more detailed analysis to the future. Consider the following $N\times N$ Higgs field in the holomorphic gauge
\begin{equation}\label{eq:holoHiggs}
\Phi^{\textrm h}=\left(\begin{array}{ccccc}0&1&0&\ldots&0\\ 0&0&1&\ldots&0\\ \vdots&\ddots&\ddots&\ddots\\ 0&0&0&\ldots&1 \\ p(v,w)&0&0&\ldots&0 \end{array}\right)\,,
\end{equation}
where $p(v,w)$ is a generic holomorphic polynomial of the worldvolume coordinates of the flat stack of $N$ D7-branes, placed at $z=0$. On one hand, at the locus where $p(v,w)$ vanishes, the dual description of this T-brane must have the same branch structure of the one studied in Section \ref{ssec:8example}: $N$ D7 sheets, each located at a different $N^{\rm th}$ root of unity in the complex $z$-plane. On the other hand, in the open set where $p(v,w)\neq0$, the matrix \eqref{eq:holoHiggs} can be diagonalized to one having all the $N$ roots of $p(v,w)$ on the diagonal.\footnote{Note that this diagonalization is achieved by a \emph{non-unitary} transformation of $SL(N,\mathbb{C})$, and hence it changes the non-holomorphic data of the T-brane configuration.} In this new holomorphic frame, this configuration has a straightforward dual description in terms of $N$ D7-branes located at $z=p(v,w)^{1/N}e^{2\pi ik/N}$, where $k=1,\ldots,N$ labels the sheets.
Hence, when moving around the $p=0$ locus, the D7-branes undergo a cyclic permutation of order $N$. This strongly suggests that the Abelian embedding dual to the T-brane \eqref{eq:holoHiggs} must be such that the $N$ sheets, which we found at $p=0$, when moving off this locus and looping around it, undergo the same type of cyclic monodromy. It would be very interesting to verify this intuition by following the prescriptions outlined in this paper in order to explicitly derive the Abelian dual of \eqref{eq:holoHiggs}.

In this paper we have restricted our analysis to T-branes in flat space. It would be very interesting to extend our duality to compact, curved T-brane configurations. Extrapolating the results presented here, we can imagine two possibilities. First, the dual Abelian brane may carry a different curvature with respect to the original T-brane, the difference of the two curvatures being related to the (non-primitive) flux data of the T-brane. Second, the dual Abelian brane may carry a non-trivial flux, which takes into account the information about the curvature of the original T-brane. Similarly, we have only considered four-supercharge T-branes with traceless flux, whereas supersymmetry only requires the trace of the flux to be a primitive two-form. We expect that allowing such extra flux on the non-Abelian side will require the dual Abelian brane to carry flux too. We hope to come back to these interesting questions in a future publication.

\vspace{.7cm}

\bigskip

\centerline{\bf \large Acknowledgments}

\bigskip

\noindent We would like to thank F.~Marchesano for useful discussions and R.~Valandro for spotting a typo in the first version. RS is grateful to the string theory group at IPhT/CEA-Saclay for kind hospitality at various stages of this project. The work of IB is supported in part by the ANR grant Black-dS-String ANR-16-CE31-0004-01 and by the John Templeton Foundation grant 61169. The work of JB is supported by MIUR-PRIN contract 2015MP2CX4002 ``Non-perturbative aspects of gauge theories and strings''. The work of RS is supported by the program Rita Levi Montalcini for young researchers (D.M. n. 975, 29/12/2014).  The work of GZ was supported by the HKRGC grants HUKST4/CRF/13G and 1630441, and the NSF CAREER grant PHY-1756996.





\begin{thebibliography}{10}






 \bibitem{Cecotti:2010bp}
	S.~Cecotti, C.~Cordova, J.~J.~Heckman and C.~Vafa,
	{\em ``T-Branes and Monodromy,''}
	JHEP {\bf 1107} (2011) 030
	[arXiv:1010.5780 [hep-th]].

\bibitem{Donagi:2003hh}
  R.~Donagi, S.~Katz and E.~Sharpe,
  {\em ``Spectra of D-branes with higgs vevs,''}
  Adv.\ Theor.\ Math.\ Phys.\  {\bf 8}, no. 5, 813 (2004)
  [hep-th/0309270].


 \bibitem{Hayashi:2009bt}
  H.~Hayashi, T.~Kawano, Y.~Tsuchiya and T.~Watari,
  {\em ``Flavor Structure in F-theory Compactifications,''}
  JHEP {\bf 1008}, 036 (2010)
  [arXiv:0910.2762 [hep-th]].

\bibitem{Chiou:2011js}
C.-C. Chiou, A.~E. Faraggi, R.~Tatar and W.~Walters,  {\em {T-branes and Yukawa
  Couplings}}, JHEP {\bf 05} (2011) 023
[\href{http://www.arXiv.org/abs/1101.2455}{{\tt 1101.2455}}].



\bibitem{Donagi:2011jy}
  R.~Donagi and M.~Wijnholt,
  {\em ``Gluing Branes, I,''}
  JHEP {\bf 1305}, 068 (2013)
  [arXiv:1104.2610 [hep-th]].




\bibitem{Donagi:2011dv}
R.~Donagi and M.~Wijnholt,  {\em {Gluing Branes II: Flavour Physics and String Duality}}, JHEP {\bf 05} (2013) 092
[\href{http://www.arXiv.org/abs/1112.4854}{{\tt 1112.4854}}].

\bibitem{Font:2013ida}
  A.~Font, F.~Marchesano, D.~Regalado and G.~Zoccarato,
  {\em ``Up-type quark masses in SU(5) F-theory models,''}
  JHEP {\bf 1311} (2013) 125
  [arXiv:1307.8089 [hep-th]].

\bibitem{Anderson:2013rka}
L.~B. Anderson, J.~J. Heckman and S.~Katz,  {\em {T-Branes and Geometry}}, JHEP
  {\bf 05} (2014) 080
[\href{http://www.arXiv.org/abs/1310.1931}{{\tt 1310.1931}}].


\bibitem{DelZotto:2014hpa}
M.~Del~Zotto, J.~J. Heckman, A.~Tomasiello and C.~Vafa,  {\em {6d Conformal
  Matter}}, JHEP {\bf 02} (2015) 054
[\href{http://www.arXiv.org/abs/1407.6359}{{\tt 1407.6359}}].

\bibitem{Collinucci:2014qfa}
	A.~Collinucci and R.~Savelli,
	{\em ``T-branes as branes within branes,''}
	JHEP {\bf 1509} (2015) 161
	[arXiv:1410.4178 [hep-th]].

\bibitem{Collinucci:2014taa}
A.~Collinucci and R.~Savelli,  {\em {F-theory on singular spaces}}, JHEP {\bf 09} (2015) 100
[\href{http://www.arXiv.org/abs/1410.4867}{{\tt 1410.4867}}].


\bibitem{Marchesano:2015dfa}
  F.~Marchesano, D.~Regalado and G.~Zoccarato,
  {\em ``Yukawa hierarchies at the point of E$_{8}$ in F-theory,''}
  JHEP {\bf 1504} (2015) 179
  [arXiv:1503.02683 [hep-th]].


\bibitem{Cicoli:2015ylx}
M.~Cicoli, F.~Quevedo and R.~Valandro,  {\em {De Sitter from T-branes}},
JHEP {\bf 03} (2016) 141
[\href{http://www.arXiv.org/abs/1512.04558}{{\tt 1512.04558}}].



\bibitem{Carta:2015eoh}
F.~Carta, F.~Marchesano and G.~Zoccarato,  {\em {Fitting fermion masses and
  mixings in F-theory GUTs}}, JHEP {\bf 03} (2016) 126
[\href{http://www.arXiv.org/abs/1512.04846}{{\tt 1512.04846}}].



\bibitem{Heckman:2016ssk}
  J.~J.~Heckman, T.~Rudelius and A.~Tomasiello,
  {\em ``6D RG Flows and Nilpotent Hierarchies,''}
  JHEP {\bf 1607}, 082 (2016)
  [arXiv:1601.04078 [hep-th]].

\bibitem{Collinucci:2016hpz}
A.~Collinucci, S.~Giacomelli, R.~Savelli and R.~Valandro,
{\em {T-branes through 3d mirror symmetry}},
  JHEP {\bf 1607}, 093 (2016)
  [arXiv:1603.00062 [hep-th]].

\bibitem{Bena:2016oqr}
  I.~Bena, J.~Blåbäck, R.~Minasian and R.~Savelli,
  {\em ``There and back again: A T-brane's tale,''}
  JHEP {\bf 1611} (2016) 179
  [arXiv:1608.01221 [hep-th]].

\bibitem{Marchesano:2016cqg}
  F.~Marchesano and S.~Schwieger,
  {\em ``T-branes and $\alpha'$-corrections,''}
  JHEP {\bf 1611} (2016) 123
  [arXiv:1609.02799 [hep-th]].

 \bibitem{Mekareeya:2016yal}
  N.~Mekareeya, T.~Rudelius and A.~Tomasiello,
  {\em ``T-branes, Anomalies and Moduli Spaces in 6D SCFTs,''}
  arXiv:1612.06399 [hep-th].

 \bibitem{Ashfaque:2017iog}
  J.~M.~Ashfaque,
  {\em ``Monodromic T-Branes And The $SO(10)_{GUT}$,''}
  arXiv:1701.05896 [hep-th].

 \bibitem{Anderson:2017rpr}
  L.~B.~Anderson, J.~J.~Heckman, S.~Katz and L.~Schaposnik,
  {\em ``T-Branes at the Limits of Geometry,''}
  arXiv:1702.06137 [hep-th].


\bibitem{Bena:2017jhm}
  I.~Bena, J.~Blåbäck and R.~Savelli,
  {\em ``T-branes and Matrix Models,''}
  JHEP {\bf 1706} (2017) 009
  [arXiv:1703.06106 [hep-th]].

\bibitem{Collinucci:2017bwv}
  A.~Collinucci, S.~Giacomelli and R.~Valandro,
  {\em ``T-branes, monopoles and S-duality,''}
  arXiv:1703.09238 [hep-th].

\bibitem{Cicoli:2017shd}
  M.~Cicoli, I.~Garcia-Etxebarria, C.~Mayrhofer, F.~Quevedo, P.~Shukla and R.~Valandro,
  {\em ``Global Orientifolded Quivers with Inflation,''}
 JHEP {\bf 1711}, 134 (2017)
  [arXiv:1706.06128 [hep-th]].


  \bibitem{Marchesano:2017kke}
  F.~Marchesano, R.~Savelli and S.~Schwieger,
  {\em ``Compact T-branes,''}
  JHEP {\bf 1709}, 132 (2017)
  [arXiv:1707.03797 [hep-th]].

\bibitem{Anderson:2017zfm}
  L.~B.~Anderson, M.~Esole, L.~Fredrickson and L.~P.~Schaposnik,
  {\em ``Singular Geometry and Higgs Bundles in String Theory,''}
  SIGMA {\bf 14}, 037 (2018)
  [arXiv:1710.08453 [math.DG]].

\bibitem{Apruzzi:2018oge}
  F.~Apruzzi, J.~J.~Heckman, D.~R.~Morrison and L.~Tizzano,
  {\em ``4D Gauge Theories with Conformal Matter,''}
  JHEP {\bf 1809}, 088 (2018)
  [arXiv:1803.00582 [hep-th]].

  \bibitem{Cvetic:2018xaq}
  M.~Cveti\v c, J.~J.~Heckman and L.~Lin,
  {\em ``Towards Exotic Matter and Discrete Non-Abelian Symmetries in F-theory,''}
  JHEP {\bf 1811}, 001 (2018)
  [arXiv:1806.10594 [hep-th]].

  \bibitem{Heckman:2018pqx}
  J.~J.~Heckman, T.~Rudelius and A.~Tomasiello,
  {\em ``Fission, Fusion, and 6D RG Flows,''}
  arXiv:1807.10274 [hep-th].

  \bibitem{Apruzzi:2018xkw}
  F.~Apruzzi, F.~Hassler, J.~J.~Heckman and T.~B.~Rochais,
  {\em ``Nilpotent Networks and 4D RG Flows,''}
  arXiv:1808.10439 [hep-th].

  \bibitem{Carta:2018qke}
  F.~Carta, S.~Giacomelli and R.~Savelli,
  {\em ``SUSY enhancement from T-branes,''}
  JHEP {\bf 1812}, 127 (2018)
  [arXiv:1809.04906 [hep-th]].


\bibitem{Marchesano:2019azf}
  F.~Marchesano, R.~Savelli and S.~Schwieger,
  ``T-branes and defects,''
  JHEP {\bf 1904} (2019) 110
  [arXiv:1902.04108 [hep-th]].


\bibitem{Myers:1999ps}
  R.~C.~Myers,
  ``Dielectric branes,''
  JHEP {\bf 9912} (1999) 022
  [hep-th/9910053].

\bibitem{Constable:1999ac}
  N.~R.~Constable, R.~C.~Myers and O.~Tafjord,
  ``The Noncommutative bion core,''
  Phys.\ Rev.\ D {\bf 61} (2000) 106009
  [hep-th/9911136].



 \bibitem{Hitchin:1986vp}
  N.~J.~Hitchin,
  {\em ``The Selfduality equations on a Riemann surface,''}
  Proc.\ Lond.\ Math.\ Soc.\  {\bf 55}, 59 (1987).

\bibitem{Tseytlin:1997csa}
  A.~A.~Tseytlin,
  ``On nonAbelian generalization of Born-Infeld action in string theory,''
  Nucl.\ Phys.\ B {\bf 501} (1997) 41
  [hep-th/9701125].


\bibitem{Hashimoto:1997gm}
  A.~Hashimoto and W.~Taylor,
  ``Fluctuation spectra of tilted and intersecting D-branes from the Born-Infeld action,''
  Nucl.\ Phys.\ B {\bf 503} (1997) 193
  [hep-th/9703217].



\bibitem{Bain:1999hu}
  P.~Bain,
  ``On the nonAbelian Born-Infeld action,''
  hep-th/9909154.

\bibitem{Myers:2003bw}
  R.~C.~Myers,
  ``NonAbelian phenomena on D branes,''
  Class.\ Quant.\ Grav.\  {\bf 20} (2003) S347
  [hep-th/0303072].


\bibitem{MAT}
I.~Bena, J.~Blåbäck, R.~Savelli and G.~Zoccarato
``Supplementary Mathematica Notebook''
\url{https://gitlab.com/johanbluecreek/T-brane_duality}


\bibitem{Green:1996dd}
  M.~B.~Green, J.~A.~Harvey and G.~W.~Moore,
  ``I-brane inflow and anomalous couplings on d-branes,''
  Class.\ Quant.\ Grav.\  {\bf 14} (1997) 47
  [hep-th/9605033].

\bibitem{Cheung:1997az}
  Y.~K.~E.~Cheung and Z.~Yin,
  ``Anomalies, branes, and currents,''
  Nucl.\ Phys.\ B {\bf 517} (1998) 69
  [hep-th/9710206].


\bibitem{Minasian:1997mm}
  R.~Minasian and G.~W.~Moore,
  ``K theory and Ramond-Ramond charge,''
  JHEP {\bf 9711} (1997) 002
  [hep-th/9710230].


\bibitem{Bachas:1999um}
  C.~P.~Bachas, P.~Bain and M.~B.~Green,
  ``Curvature terms in D-brane actions and their M theory origin,''
  JHEP {\bf 9905} (1999) 011
  [hep-th/9903210].


\bibitem{Minasian:2001na}
  R.~Minasian and A.~Tomasiello,
  ``Variations on stability,''
  Nucl.\ Phys.\ B {\bf 631} (2002) 43
  [hep-th/0104041].


\bibitem{Butti:2007aq}
  A.~Butti, D.~Forcella, L.~Martucci, R.~Minasian, M.~Petrini and A.~Zaffaroni,
  ``On the geometry and the moduli space of beta-deformed quiver gauge theories,''
  JHEP {\bf 0807} (2008) 053
  [arXiv:0712.1215 [hep-th]].


\bibitem{Marchesano:2010bs}
  F.~Marchesano, P.~McGuirk and G.~Shiu,
  ``Chiral matter wavefunctions in warped compactifications,''
  JHEP {\bf 1105} (2011) 090
  [arXiv:1012.2759 [hep-th]].




\end{thebibliography}
\end{document}